\documentclass[graybox]{svmult}

\usepackage{mathptmx}       
\usepackage{helvet}         
\usepackage{courier}        
\usepackage{type1cm}        
\usepackage{makeidx}         
\usepackage{graphicx}        
\usepackage{multicol}        
\usepackage[bottom]{footmisc}
\usepackage{enumitem}
\makeindex             

\newcommand{\tc}{T$_c$}
\newcommand{\ep}{$ep$}
\newcommand{\mg}{MgB$_2$}
\newcommand{\sh}{SH$_3$}
\newcommand{\ph}{PH$_3$}

\begin{document}

\title*{Understanding Novel Superconductors with {\em ab-initio} Calculations}
\author{Lilia Boeri}
\institute{Lilia Boeri \at Dipartimento di Fisica, Sapienza Universit\`a
  di Roma, P.le Aldo Moro 2, 00185 Roma, Italy,
\email{lilia.boeri@uniroma1.it}}

%
%
\maketitle

\abstract{
  This chapter gives an overview of the progress in the
  field of computational superconductivity.
  Following the \mg\ discovery (2001), there has been an impressive
  acceleration in the development of methods based on Density Functional Theory   to compute the  critical temperature and other physical
  properties of actual superconductors from first-principles.
  State-of-the-art {\em-ab-initio} methods have reached predictive accuracy
  for conventional (phonon-mediated) superconductors, and substantial
  progress is being made also for unconventional superconductors.
  The aim of this chapter is to give an overview of the existing computational
  methods for superconductivity, and present selected examples of material discoveries that exemplify the main advancements.}

\section{Introduction}
\label{sect:intro}


The aim of this chapter is to offer an up-to-date perspective on the field of
{\em ab-initio} superconductivity and of the related development 
of numerical methods to compute
critical temperatures and other physical 
properties of superconductors.

The material-specific aspect is what distinguishes {\em ab-initio}  (= from first principles) approaches, based on Density Functional Theory (DFT) and its extensions, from other theoretical approaches to superconductivity, which mainly focus on the general description of the phenomenon.
The typical questions addressed by computational superconductivity are:
($i$) what makes a certain compound a good (or bad) superconductor? ($ii$) How are its properties modified by external parameters, such as doping, pressure, strain?
($iii$) Is it possible to find new materials with improved superconducting properties compared to existing ones?

The most relevant parameter that defines the performance
of a superconductor for large-scale applications
is its critical temperature (\tc):
this means that addressing the questions above requires
the development of methods
accurate enough to predict the \tc\ of a superconductor, and its dependence
on external parameters. The progress in this direction, in the last
twenty years, has been impressive. 

For a large class of superconductors, i.e. conventional, phonon-mediated
ones, {\em ab-initio} methods are now so accurate that the
focus of the field is gradually shifting from the description of
existing superconductors to the design of new materials. The first
successful example  was the prediction of
high-\tc\ conventional superconductivity in \sh\ (2014).~\cite{H:DrozdovEremets_Nature2015,H:Duan_SciRep2014}

For unconventional superconductors, which comprise two of the
most studied classes of materials, the high-\tc\ cuprates~\cite{SC:Bednorz_ZPB_1986} and
Fe-based superconductors,~\cite{fesc:kamihara_JACS_2008}
{\em ab-initio} approaches are still far from being predictive, but
it is becoming more and more widely accepted that the single-particle
electronic structure determines crucial properties of these materials,
such as the symmetry of the superconducting gap and the behavior of
magnetic excitations.

The  topics and the structure of this chapter have been
specifically thought to illustrate the parallel progress
in {\em ab-initio} methods and material research
for superconductors.
I have chosen three discoveries that I consider the fundamental
milestones of this process:
($a$) The report of superconductivity in \mg\ in 2001, which has disproved the Cohen-Anderson limit for
conventional superconductors;
($b$) The discovery Fe-based superconductors, which has lead to a much deeper understanding of the interplay between electronic structure, magnetism and
superconductivity in unconventional superconductors; ($c$) The discovery of high-temperature superconductivity at Megabar pressures in \sh, which has given a spectacular demonstration the predictive power of {\em ab-initio} calculations.

Although I will give a general introduction to the theory of superconductivity
and briefly describe the most recent advancements in {\em ab-initio}
methods, methodological developments are not the main topic of
this chapter: I refer the interested
reader to excellent reviews in literature for a detailed discussion.~\cite{DFT:giustino_RMP_2017,DFT:Sanna_SCDFT_2017,DFT:Sanna_Eliashberg_JPSJ_2018}
For space reasons, I am also forced to leave out some currently
very active
directions of superconductivity research, 
such as  topological superconductivity,~\cite{SC:Sato_topological_review} superconductivity in 2D transition metal dichalcogenides,~\cite{SC:KLEMM_2D}
artificial superlattices\cite{SC:Mannhart_review} and 
other more traditional topics, such as cuprates and other oxides,~\cite{SC:CHU_cuprates} fullerenes, \cite{SC:Gunnarsson_RMP_1997}, layered halonitrides,~\cite{SC:KASAHARA_NCL}, etc. 

On the other hand, have included at the end a short perspective
describing possible routes to high-\tc\ superconductivity
which exploit novel developments in experimental and
{\em ab-initio} techniques, since I believe that in the
next years the combination of the two may lead to the discovery of many
new superconductors.

The structure of the chapter is the following: I will start by giving
a concise historical review of the most important discoveries
in section~\ref{sect:history}.
In section~\ref{sect:theory} I will then introduce
the basic concepts of superconductivity theory, and
describe the most recent developments in {\em ab-initio} methods.
The main body of the paper is contained in Section~\ref{sect:materials},
where, using selected material examples, I will try to give
an impression of the rapid progress of the field in the understanding of
both conventional and unconventional superconductors.
Finally, in  section~\ref{sect:outlook} I will propose possible practical
routes to high-\tc\ superconductivity.

\section{A Brief History of Research in Superconductivity}
\label{sect:history}

Superconductivity was discovered more than 100 years ago when H.K. Onnes
observed that, when cooled below 4 K, mercury exhibits a vanishing resistivity.~\cite{SC:Onnes}
Perfect diamagnetism, which is the second fingerprint of a superconductor, was discovered by Meissner and Ochsenfeld around 20 years later.~\cite{SC:Meissner} 
While it was immediately clear that superconductors could have an enormous potential for applications, the low critical temperatures 
represented an insurmountable obstacle to  large-scale applications.

In addition to presenting practical problems,
superconductivity proved to be
a major challenge also for theorists:
fully microscopic theories of superconductivity  - the Bardeen-Cooper-Schrieffer (BCS) and Migdal-Eliashberg (ME) theories - were
developed only after almost fifty years after the original discovery;~\cite{Th:BCS_PR_1957,Th:Migdal_JETP_1958,Th:Eliashberg_JETP_1960,Th:Allenmit_book_1982,Th:Scalapino_book_1969,Th:Carbotte_RMP_1990}.
They describe superconductivity as due to the condensation of pairs of electrons of opposite spin and momentum (Cooper pairs), held together by an attractive {\em glue}.~\footnote{We will treat here only the case of so-called {\em boson-exchange} superconductors, and not other mechanisms, such as resonant valence bond, hole superconductivity, etc}
In conventional superconductors, the glue is provided by phonons (lattice vibrations), but other excitations such as plasmons, spin fluctuations etc can also mediate the superconducting pairing.

The understanding of the microscopic mechanism of superconductivity
did not lead to any immediate, appreciable
progress in the search for new superconductors;
this translated into a general skepticism towards theory, which is well exemplified by one of the Matthias' rules
for superconductivity ({\em stay away from theorists!}). Indeed, rather than a predictive theory, ME theory was long considered a sophisticated
phenomenological framework 
to describe existing superconductors, while the search for new materials was (unsuccesfully) guided by semi-empirical rules. Even worse, two leading
theorists used ME theory to demonstrate 
the existence of an intrinsic limit of around 25 K to the \tc\ of {\em conventional} superconductors.
Although conceptually wrong, the {\em Cohen-Anderson limit} is still cited today
as an argument against high-\tc\ superconductivity.~\cite{Th:Cohen_anderson}

The notion of an upper limit to \tc\ was first challenged by the discovery of the first unconventional superconductors, the cuprates, in 1986.~\cite{SC:Bednorz_ZPB_1986}
In contrast to conventional superconductors, which above \tc\ behave as
ordinary metals, cuprates exhibit a complex phase diagram, with many coexisting phases and physical phenomena (charge and spin density waves,
metal-insulator transitions, transport anomalies etc).
In 1987 a cuprate, YBCO, broke the liquid N$_2$ barrier, with a \tc\ of 92 K,~\cite{SC:YBCO} causing a general excitement in the media about a possible {\em superconducting revolution}; 
the highest \tc\ ever attained in this class is 156 K.~\cite{SC:cuprate_Hg1993}
Despite almost thirty years of research, and many different proposals,
a quantitative theory of
superconductivity in the cuprates is still lacking; furthermore,
their large-scale applicability is also limited due to their high brittleness and manufacturing costs.~\cite{SC:Gurevich_nphys_2011}

\begin{figure}[h!]
\includegraphics[scale=.50]{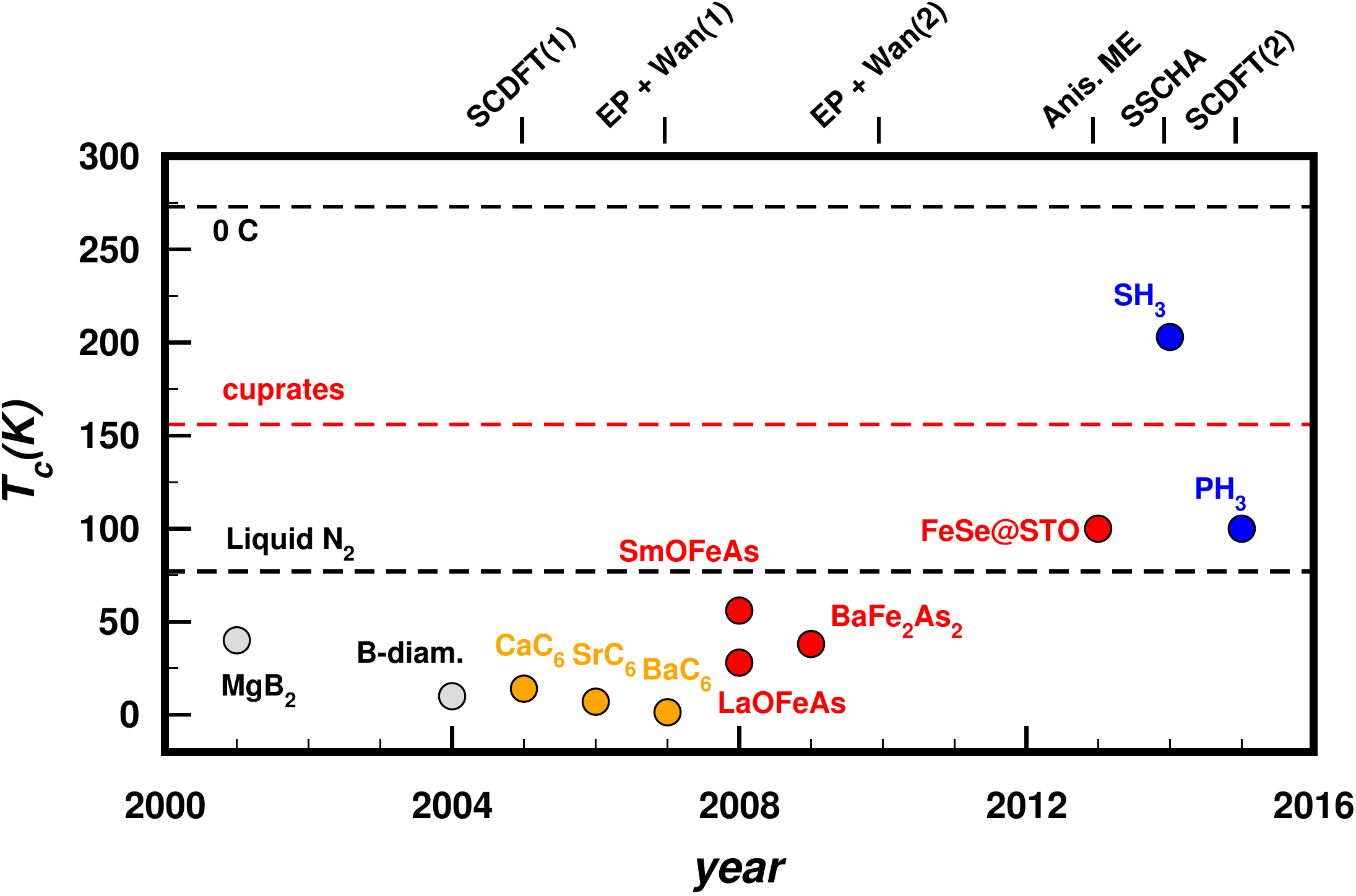}
\caption{Main Developments in the field of Superconductivity in the 21$^{st}$ century: The \tc's of the most important experimental discoveries as a function
  of year are shown as colored symbols.
  The top axis reports the most important methodological developments
in {\em ab-initio} supercoductivity:
Superconducting Density Functional Theory ({\em SCDFT(1)}),~\cite{Th:Luders_PRB_2005,Th:Marques_PRB_2005} electron-phonon interaction with Wannier functions
({\em EP+WAN(1) and EP+WAN(2)}),~\cite{DFT:giustino_RMP_2017,DFT:Giustino_PRB_2007,DFT:profeta_PRB_2010}
{\em ab-initio} anisotropic ME theory ({\em Anis. ME}),~\cite{DFT:Margine_PRB_2013,DFT:Sanna_Eliashberg_JPSJ_2018} (stochastic) self-consistent harmonic approximation ({\em SSCHA}),~\cite{DFT:Errea_PRB_2014}
{\em ab-initio} spin-fluctuations ({\em SCDFT(2)}).~\cite{DFT:Essenberger_PRB_2014,DFT:Essenberger_PRB_2016}.
}
\label{fig:history}
\end{figure}

The search for new materials took a different turn at the beginning of this century, when a \tc\ of 39$\;K$ was reported in a 
simple $s$--$p$ binary compound,
magnesium diboride (\mg).~\cite{SC:akimitsu_mgb2}
In contrast to the cuprates, \mg\ is a conventional superconductor.
In less than two years, {\em ab-initio} calculations provided a key quantitative understanding of very specific aspects of superconductivity in this material,
such as two-gap superconductivity, anharmonicity, role of magnetic and non-magnetic impurities, doping, etc. 
~\cite{SC:mgb2:pickett,SC:mgb2:kong,SC:mgb2:kortus1,SC:mgb2:kortus2,SC:mgb2_choi_2002,SC:mgb2_clean_or_dirty}
This stimulated a renewed enthusiasm in the search for new superconductors and, in parallel, the development of accurate {\em ab-initio} methods
to model them.

Indeed, in the last seventeen years superconductivity has been discovered in B-doped semiconductors,~\cite{SC:diamond:Ekimov_2004,SC:silicon:Bustarret_2006}, intercalated graphites, ~\cite{SC:weller_nphys_2005}
unconventional Fe-based superconductors ~\cite{fesc:kamihara_JACS_2008}
and, finally, high-pressure hydrides.~\cite{H:DrozdovEremets_Nature2015,H:Drozdov_PH3_arxiv2015} The \tc's as a function of the year of
their discovery are shown as colored symbols in Fig.~\ref{fig:history}.
The most important methodological developments are reported on the top axis:
Superconducting Density Functional Theory ~\cite{Th:Luders_PRB_2005,Th:Marques_PRB_2005}, electron-phonon interaction with Wannier functions~\cite{DFT:Giustino_PRB_2007,DFT:profeta_PRB_2010,DFT:giustino_RMP_2017},
{\em ab-initio} anisotropic ME theory
~\cite{DFT:Sanna_Eliashberg_JPSJ_2018,DFT:Margine_PRB_2013}, (stochastic) self-consistent harmonic approximation,~\cite{DFT:Errea_PRB_2014}, {\em ab-initio}
spin-fluctuations.~\cite{DFT:Essenberger_PRB_2014,DFT:Essenberger_PRB_2016}.

Thanks to these advancements, {\em ab-initio} calculations for conventional superconductors have now reached an accuracy which gives them
fully predictive power. Methods are being developed also to treat other types of interactions, such as plasmons and spin fluctuations,~\cite{DFT:Essenberger_PRB_2014,DFT:Essenberger_PRB_2016,DFT:Akashi_plasmon_PRL_2013}
and parameter ranges where the standard approximations of strong-coupling ME theory break down.~\cite{Th:pietronero_PRL_1995}.
Combined with the development of efficient methods for {\em ab-initio} crystal structure prediction and
the progress in synthesis and characterization
techniques,~\cite{DFT:Woodley_Catlow_nmat_2008,SC:Zhang2017_highP_review}
this opens unprecedented
possibilities for material discovery.
\section{Methods}
\label{sect:theory}

In this section I describe the methodological background of computational superconductivity. The first part introduces the main concepts behind
the microscopic theories of superconductivity, i.e. the early Bardeen-Cooper-Schrieffer (BCS) theory and the strong coupling Migdal-Eliashberg (ME) theory.
Although extremely accurate and elegant, ME theory was for a long time employed only as a semi-phenomenological theory, relying on
electronic and vibrational spectra extracted from experiments. 
Early attempts of obtaining this quantities from
Density Functional Theory date back to the early 
80's,~ but with limited success, due to
inadequate computational resources and insufficient accuracy
of methods to treat phonons and electron-phonon (\ep) interaction.~\cite{DFT:Gaspari_PRL_1972,DFT:Cohen_SC_Si_PRL_1985}
The required accuracy was only achieved with Density Functional
Perturbation Theory (DFPT),~\cite{DFT:Savrasov_PRB_1996,DFT:Baroni_RMP_2001} and 
recently substantially increased with Wannier-function interpolation methods.~\cite{DFT:Giustino_PRB_2007,DFT:profeta_PRB_2010,DFT:Marzari_RMP_2012}

In a very influential paper, already in 1996, Savrasov and Savrasov demonstrated that linear response calculations combined with Migdal-Eliashberg theory could reproduce the critical temperatures and other properties of elemental metals.~\cite{DFT:Savrasov_PRB_1996}
However, since at that time the \tc's of known conventional
superconductors were much smaller than those of the cuprates,
this result  was erroneously perceived as of limited importance.

The report of superconductivity in \mg\ in 2001
gave a strong impulse to the development of
{\em ab-initio} methods for superconductors,
which resulted in two parallel lines of research:
 {\em ab-initio} Migdal-Eliashberg theory and Superconducting Density Functional Theory, described in Sect.~\ref{sect:theory:DFT}.
In their fully anisotropic versions, including screened Coulomb interactions,
they have a comparable accuracy of 5-10$\%$ on the critical temperature, gap, etc.~\cite{DFT:Sanna_SCDFT_2017,DFT:Sanna_Eliashberg_JPSJ_2018,DFT:Margine_PRB_2013} This gives them fully predictive power, and, combined
with methods for crystal structure prediction,
offers the unprecedented possibility of designing superconductors
{\em ab-initio}, overcoming the practical limitations of experiments.

\subsection{A Short Compendium of Superconductivity Theory}
\label{sect:theory:general}
The first fully microscopic theory of Superconductivity was formulated by
Bardeen, Cooper and Schrieffer in 1957, and is known as BCS theory.~\cite{Th:BCS_PR_1957}

BCS theory describes the transition of superconductors from an ordinary metallic state ({\em normal state})
to a new state, characterised by vanishing resistivity and perfect diamagnetism.
In this {\em superconducting state} the electronic spectrum develops a gap $\Delta$ around
the Fermi level, which is maximum at zero temperature and vanishes at the critical temperature \tc.
The critical temperature exhibits an {\em isotope effect},
i.e. \tc\ increases (or decreases) upon partial replacement of an element with a ligther (heavier) isotope.
Isotope effects are also measured for other characteristic properties of superconductors (gap, specific heat, etc.).

BCS theory reconciles all the above
experimental observations in a consistent framework,
based on three key concepts:

($i$) A Fermi sea
of electrons in the presence of an attractive interaction is unstable 
towards the formation of a pair of electrons with opposite spin and momentum
({\bf Cooper pair}), which effectively behaves as a boson.
In a superconductor, below \tc,  a small, but macroscopic fraction of electrons, of order $\Delta/E_F \simeq 10^{-3}$,
forms Cooper pairs --
this is sometimes referred to as the {\em condensate fraction} of a
superconductor.

($ii$) A variational many-body wavefunction for the electrons
is constructed from a superposition
of ordinary single-particle states and Coooper pairs ({\bf BCS wavefunction}).
The existence of a  condensate fraction leads to the appearance of a gap in the electronic spectrum
$\varepsilon_{\mathbf{k}}$, which satisfies the self-consistent equation:
\begin{equation}
\Delta_{\mathbf{k}}=-\frac{1}{2}\sum_{\mathbf{k},\mathbf{k}'}\frac{V_{\mathbf{k},\mathbf{k}'}\Delta_{\mathbf{k}'}}{\sqrt{\varepsilon^2_{\mathbf{k}} + \Delta^2_{\mathbf{k}}}}\cdot \tanh\left(\frac{\sqrt{\varepsilon^2_{\mathbf{k}}+\Delta^2_{\mathbf{k}}}}{2T} \right).
\label{eq:BCSgap}
\end{equation}

$iii$) Using a simple model for the electron-electron interaction 
$V_{\mathbf{k},\mathbf{k}'}$, the so-called {\bf BCS potential},
which is attractive only if the two electrons
with wavevector $\mathbf{k},\mathbf{k}'$ both
lie in a small energy shell  $\omega_D$ around the Fermi energy
$E_F$, Eq.~\ref{eq:BCSgap} can be solved analytically, and the gap
and \tc\ are given by:
\begin{equation}
\Delta(T=0) \simeq 2 \omega_D \exp\left(-\frac{1}{ N(E_F) V}\right), \hspace{0.5cm} k_b T_c=1.13  \omega_D \exp\left(-\frac{1}{ N(E_F) V}\right),
\label{eq:BCSTc}
\end{equation}

The original idea of Bardeen, Cooper and Schrieffer is that the attractive interaction $V$ between electrons is mediated by lattice vibrations (phonons);
In this case,
$\omega_D$ is a representative phonon energy scale,
such as the Debye frequency; $N(E_F)$ is the electronic Density of
States at the Fermi level.

One of the first successes of BCS theory has been the explanation of the isotope effect on \tc:  
$\alpha_{T_c}=-\frac{d \ln (T_c)}{d \ln (M)}=0.5$, where $M$ is the ionic mass,
as well as the prediction of several {\em magic ratios}, 
satisfied in most elemental superconductors: the most famous is probably the ratio  $\frac{2 \Delta(0)}{k_B T_c}=3.53$.

However, BCS theory is valid only at weak coupling ($\lambda=N(E_F)V < 0.2-0.3$) and instantaneous interactions;
these assumptions are not verified in many  conventional superconductors,
where the actual values of \tc, isotope effect, magic ratios etc are spectacularly different from the BCS predictions.~\cite{Th:Marsiglio_Carbotte_1986}.

A quantitative description of the strong-coupling, retarded regime
is given by the many-body ME theory,
based on a set of self-consistent coupled diagrammatic equations for the electronic and bosonic propagators.~\cite{Th:Allenmit_book_1982,Th:Scalapino_book_1969,Th:Carbotte_RMP_1990,DFT:Margine_PRB_2013}.
The bosons that mediate the superconducting pairing can be
phonons or other excitations of the crystal, 
such as plasmons or spin fluctuations.~\cite{Th:Berk_Schrieffer_PRL_1966} 
Below the critical temperature, electrons are described by
a normal and an {\em anomalous} electronic propagator,
the latter accounting for Cooper pairs.
ME equations are then obtained from the Dyson's equations for
the normal and anomalous propagators;
in their most commonly-used, $T$-dependent form they can be written as:
\begin{eqnarray}
  Z(\mathbf{k},i\omega_n)&=&1+\frac{\pi T}{\omega_n}\sum_{\mathbf{k}'n'}\frac{\delta(\varepsilon_{\mathbf{k}'})}{N(E_F)}\frac{\omega_{n'}}{\sqrt{\omega_{n'}^2+\Delta^2(\mathbf{k},i\omega_{n'})}}\lambda(\mathbf{k},\mathbf{k}',n-n')
\label{eq:ME1} 
\\
Z(\mathbf{k},i\omega_n)\Delta(\mathbf{k},i\omega_n)&=&\pi T \sum_{\mathbf{k}'n'}\frac{\delta(\varepsilon_{\mathbf{k}'})}{N(E_F)}\frac{\Delta(\mathbf{k}',i\omega_{n'})}{\sqrt{\omega_{n'}^2+\Delta^2(\mathbf{k},i\omega_{n'})^2}} \times
\nonumber
\\
&\times&
\left[ \lambda(\mathbf{k},\mathbf{k}',n-n')-\mu(\mathbf{k}-\mathbf{k}'))\right]
\label{eq:ME2}
\end{eqnarray}
where $Z(\mathbf{k},i\omega_n)$ and $Z(\mathbf{k},i\omega_n) \Delta(\mathbf{k},i\omega_n)$ are the self-energy of the normal and anomalous electronic propagators, respectively; 
$i\omega_n=i(2n \pi T+1)$ are Matsubara frequencies, $\mathbf{k},\mathbf{k}'$ are the electronic momenta; the $\delta$ function
restricts the sum over {\bf k}' only to electronic states at the Fermi level.
The electrons interact through a retarded attractive interaction $\lambda(\mathbf{k},\mathbf{k}',n-n')$, and an instantaneous Coulomb repulsion 
$\mu (\mathbf{k}-\mathbf{k}')$.
The interaction $\lambda(\mathbf{k},\mathbf{k}',n-n')$ is usually expressed in terms of an electron-boson spectral function $\alpha^2 F(\mathbf{k},\mathbf{k}',\omega)$ as :
\begin{equation}
\lambda(\mathbf{k},\mathbf{k}',n-n') =  \int_0^{\infty}d \omega \frac{2 \omega}
      {\left(\omega_n - \omega_{n'}\right)^2+\omega^2}
  \alpha^2 F(\mathbf{k},\mathbf{k}',\omega)
\label{eq:alphaaniso}  
\end{equation}

Eqs.~(\ref{eq:ME1}-\ref{eq:ME2})
can be solved numerically to obtain the gap, and other thermodynamic quantities.
It is very common, and in most cases sufficiently accurate, to approximate
the more general form with an isotropic version,
replacing the sums on the electronic momenta ($\mathbf{k},\mathbf{k}'$) with
Fermi surface averages.
If one is only interested in the \tc,
there are excellent approximation formulas; a very popular
choice for phonon-mediated superconductors is the 
Mc-Millan-Allen-Dynes expression:~\cite{Th:AllenDynes_PRB_1975}
\begin{equation}
\label{eq:McMillan}
  T_c=\frac{\omega_{\log}}{1.2 k_B}\exp\left[-\frac{1.04(1+\lambda)}{\lambda-\mu^{*}(1+0.62\lambda)}\right]~,
\end{equation}
where $\lambda=2 \int d\omega \frac{\alpha^2 F(\omega)}{\omega}$ and $\omega_{\log}=\exp\left[\frac{2}{\lambda}\int \frac{d\omega}{\omega} \alpha^2 F(\omega)\ln(\omega) \right]$ are
the $ep$ coupling constant and logarithmic averaged phonon frequency, respectively; 
 $\mu^*$ is the so-called Morel-Anderson pseudopotential,  obtained by
screening the full Coulomb potential up to a characteristic cut-off energy.~\cite{Th:Morel_Anderson_1962}


\subsection{{\em Ab-initio} methods:}
\label{sect:theory:DFT}

The two methods described in this section, {\em ab-initio} anisotropic
Midgal-Eliashberg Theory (DFT-ME in the following)~\cite{DFT:Margine_PRB_2013,DFT:Sanna_Eliashberg_JPSJ_2018}  and SuperConducting Density Functional Theory (SCDFT),~\cite{Th:Oliveira_PRL_1988,Th:Luders_PRB_2005,Th:Marques_PRB_2005}
represent a fundamental step forward in the study of actual superconductors,
because they permit to obtain a full characterization of the normal
and superconducting state of a system from the sole knowledge of
the chemical composition and crystal structure.
Although there are fundamental and practical differences between the
two, both methods rely crucially on the ability of
DFT of providing accurate electronic and bosonic spectra for most
materials  at an affordable computational cost.~\cite{DFT:Jones_RMP_1989,DFT:Baroni_RMP_2001}

The basic assumptions are the following:
\begin{enumerate}
\item {\em Electronic quasi-particles} appearing in Eqs.~(\ref{eq:ME1}-\ref{eq:ME2}) are replaced by the Kohn-Sham quasi-particles.
\item {\em Bosonic} excitation energies and electron-boson spectral functions are obtained from Density Functional Perturbation Theory (DFPT).~\cite{DFT:Baroni_RMP_2001}

For {\em phonons}, the spectral function is:
\begin{equation}
  \alpha^2 F_{ph} (\mathbf{k},\mathbf{k}',\omega)=
  N(E_F)\sum_{\nu} g_{\mathbf{k}\mathbf{k}',\nu}
 \delta (\omega - \omega_{\mathbf{k}-\mathbf{k}',\nu}), 
\label{eq:alpha_phonon}
\end{equation}
  where $g_{\mathbf{k},\mathbf{k'}}=\left< \mathbf{k}' \vert \delta V_{scf}^{\nu, \mathbf{k}'-\mathbf{k}} \vert \mathbf{k} \right>$ is the \ep\ matrix element for the mode $\nu$;
 and  $\delta V^{\nu,\mathbf{q}}_{scf}$  is the variation of the Kohn-Sham self-consistent potential due to an infinitesimal
 displacement along the eigenvector of the phonon mode $\nu$ with wave-vector
 $\mathbf{q}=\mathbf{k}'-\mathbf{k}$.
  
 For {\em spin fluctuations}, the spectral function is proportional to the imaginary part of the longitudinal interacting spin susceptibility
 $\chi_{zz}(\mathbf{q}=\mathbf{k}-\mathbf{k}',\omega)$\cite{DFT:vignale_PRB_1985}, which, using linear response within the time-dependent-Density-Functional-Theory (TDDFT) framework,~\cite{DFT:TDDFT_grosS_PRL1984} can be written as:
    \begin{equation}
\chi_{zz} (\mathbf{q},\omega)=\frac{\chi^{KS}(\mathbf{q},\omega)}{1-f_{xc}(\mathbf{q},\omega)\chi^{ks}(\mathbf{q},\omega)}, 
    \label{eq:alpha_SF}
  \end{equation}
    where $\chi^{KS}(\mathbf{q},\omega)$ is the Kohn-Sham susceptibility, and $f_{xc}(\mathbf{q},\omega)$
    is the exchange and correlation kernel.~\cite{DFT:Essenberger_PRB_2012}

  \item The {\em Coulomb potential} $\mu(\mathbf{k}-\mathbf{k}')$ which, in most
    empirical approaches, is treated as an adjustable parameter within the
    Morel-Anderson approximation,~\cite{Th:Morel_Anderson_1962}
    is treated fully {\em ab-initio} screening
    the bare Coulomb potential within RPA.
    Substantial deviations from typical values of $\mu^*=0.1-0.15$
    are found in strongly anisotropic systems such as \mg\
and layered superconductors; in some cases, such as alkali metals at high pressure, the effect of Coulomb interactions is even stronger, giving rise to plasmonic effects. ~\cite{DFT:Akashi_plasmon_PRL_2013}
  \end{enumerate}

Once the spectra of the quasi-particles and the interactions between them are known from first-principles, DFT-ME or SCDFT can be applied
to describe the superconducting state. DFT-ME theory amounts to solving the fully anisotropic ME equations, for electronic and bosonic spectra computed in DFT;
The current implementations solve the equations in Matsubara frequencies, and use Pad\'e approximants to continue them to real space. The obvious advantage of this method is that all quantities have an immediate physical interpretation trhough many-body theory.

SCDFT is a fundamentally different approach, that generalizes the original
Hohenberg-Kohn idea of one-to-one correspondence between ground-state density and potential,~\cite{DFT:HK} introducing two additional densities (and potentials) for the ionic system $\Gamma(\mathbf{R}_i) (V_{ext}(\mathbf{R}_i))$ and
 the superconducting electrons $\chi(\mathbf{r},\mathbf{r}') (\Delta (\mathbf{r},\mathbf{r}'))$, and finding the values that minimize a suitable energy functional.
This permits to derive a gap equation which is analogous to the BCS one,
but instead of an empirical potential contains a {\em kernel} with all the relevant physical information on the system.
I refer the reader to the original references for a full
derivation,~\cite{Th:Luders_PRB_2005,Th:Marques_PRB_2005,Th:Oliveira_PRL_1988}
and to Ref.~\cite{DFT:Sanna_SCDFT_2017} for an excellent pedagogical introduction.
 
SCDFT equations are more easily solvable on a computer than fully anisotropic
ME equations because they do not require expensive sums over Matsubara
frequencies; however, the interpretation of many physical quantities, including the frequency-dependence of the gap, is not equally transparent and straightforward. Another intrinsic limitation is that, like in all DFT-like methods, the quality of the results depends strongly on the quality of the functional.

The latest-developed functionals yield results with an {\em accuracy}
comparable to that of the best DFT-ME calculations, which for most conventional superconductors is between 5 and 10$\%$ of the critical temperature.
The most severe source of inaccuracy in DFT calculations for superconductors 
is usually an underconverged integration in reciprocal
space in Eqs.~(\ref{eq:ME1}-\ref{eq:ME2}), an issue that has
considerably improved thanks to the use of Wannier interpolation techniques.

Achieving quantitative accuracy for conventional pairing also encouraged
to address {\em ab-initio} effects, which are often disregarded even in model approaches, such as anharmonicity, vertex corrections, and zero-point
effects.~\cite{Th:PGS_PRB_1995,mine:Boeri_PRB_2005,SC:lazzeri_mgb2_2003}
  
These turned out to be relevant for a wide variety of compounds, particularly
for the newly-discovered superconducting hydrides, where the energy scales
of phonons and electrons are comparable.~\cite{DFT:Errea_PRB_2014,H:Sano_SH3_PRB_2016,H:errea_nature_2016}
I refer the reader to the relevant references for an in-depth discussion.

For unconventional superconductors, on the other hand, the most severe source of inaccuracy is intrinisc and is the possible divergence in the spin-fluctuation propagator, which completely destroys the predictive power of DFT approaches for currently available functionals.
Moreover, most unconventional superconductors suffer from the lack of
accuracy of DFT in strongly correlated systems.~\cite{DFT:Yin_PRX2011}

\subsection{Developments in related fields: {\em ab-initio} Material Design}
\label{sect:theory:crystal}
The term {\em ab-initio Material design} indicates the combination of methods for {\em ab-initio} crystal structure prediction and thermodynamics to
predict the behaviour of materials at arbitrary conditions of pressure and temperature, knowing only the initial chemical composition of the system.
The development of these methods represents a substantial step forward in computational condensed matter research, as it overcomes one of its biggest limitations,
exemplified by the {\em Maddox Paradox} (1988):
{\em  'One of the continuing scandals in the physical sciences is that it remains in general impossible to predict the structure of even the simplest crystalline solids from a knowledge of their chemical composition'}.

The basic working principle of {\bf {\em ab-initio} crystal structure prediction} is quite simple.
Predicting the crystal structure of a material for a given regime of chemical composition and pressure amounts to finding the global minimum of a complicated landscape,
generated by the {\em ab-initio} total energies (or enthalpies) of all possible structures.
The number of possible configurations for a typical problem is so large
that a purely enumerative approach is unfeasible; in the last years, several
methods have been devised to make the problem computationally manageable,
such as {\em ab-initio} random structure search, minima hopping, metadynamics, genetic algorithms, particle swarm algorithms, etc.~\cite{DFT:Woodley_Catlow_nmat_2008}

\begin{figure}[h!]
\includegraphics[width=12cm,clip,angle=0]{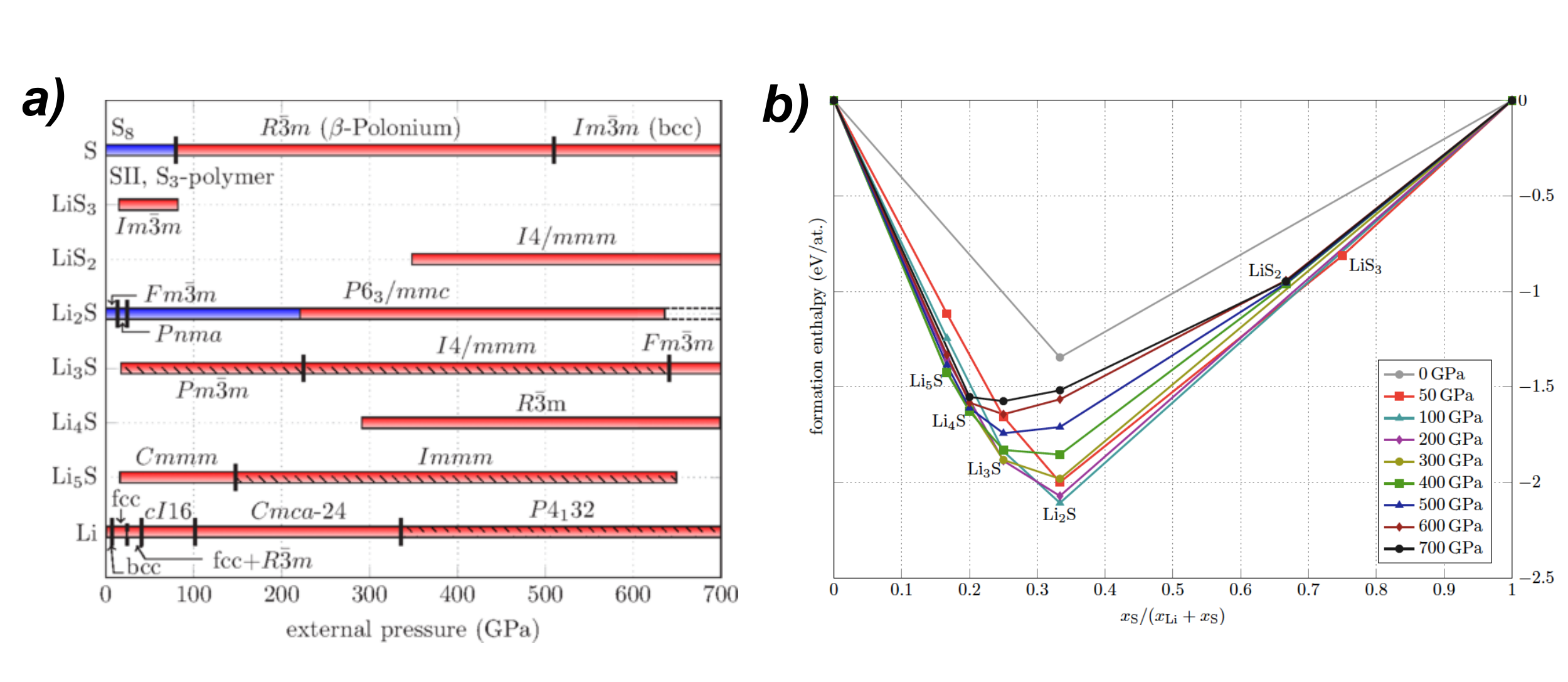}
\caption{Basic steps of {\em ab-initio} materials prediction,
  readapted from Ref.~\cite{mine:Kokail_PRB_2016}: the complete phase diagram of a binary alloy as a function of pressure ($a$) can be constructed combining several convex hull constructions at different pressures ($b$) ({\em see text}).}
\label{fig:crystal}
\end{figure}

Once the most favorable crystal structure for a given composition and pressure is known, the {\em ab-initio} Calphad (CALculation of PHAse Diagrams) approach permits to predict accurate {\bf phase diagrams},~\cite{DFT:calphad_book}
as illustrated in Fig.~\ref{fig:crystal}.
The binary Li-S phase diagram in panel $(a)$ shows the stability ranges of different Li-S compositions and has been constructed repeating several {\em convex hull} calculations at different pressures.

The {\em convex hull} construction is shown in panel ($b$).
The points represent the lowest-energy structure predicted by an evolutionary search for a given composition; for a binary phase with composition $A_xB_y$, the formation enthalpy $\Delta H$ is defined as: 
$
\Delta H = H (A_xB_y) - \left[x H(A) + y H(B)\right].
$
If this quantity is negative, the phase $A_xB_y$ is stable with respect to elemental decomposition; if it is positive, the phase is highly (or weakly) metastable, i.e., if formed, it will decompose into its elemental constituents in finite time. However, the decomposition into the two elements is often not the relevant one, as a compound could decompose into other phases, preserving the correct stoichiometry. To estimate all possible decompositions for the binary system, one constructs the most convex curve connecting the formation enthalpy of all known phases for a given stoichiometry. Points on this convex hull represent stable compositions into which phases which lie above the convex hull will decompose into given a sufficient interval of time. The figure also shows that, as pressure increases, the diversity of the phase diagram increases, i.e. off-stoichiometry compositions
become possible ({\em forbidden chemistry}).
The convex hull construction can be easily extended to multinary systems (Gibbs diagrams) and finite temperatures including entropic effects.

\section{Materials:}
\label{sect:materials}

The aim of this section is to illustrate how, within a little bit more than a decade, an increased understanding of material-specific aspects of superconductivity gained from {\em ab-initio} calculations has permitted to replace empirical
rules to search for new superconductors with quantitative strategies.

I will start with a general discussion that determine the \tc\ of  conventional superconductors (Sect.~\ref{sect:mat:conventional}),
introducing the concepts of
{\em dormant} \ep\ interactions and lattice instabilities,
and showing how these can be used to interpret both the old empirical
knowledge (Matthias' rules and Cohen-Anderson limit) and the latest
experimental discoveries.~\cite{DFT:Savrasov_PRB_1996,mine:Boeri_PRB_2001,mine:Boeri_PRB_2007,mine:Boeri_PRL_2008,SC:Profeta_Graphene_Nat2012}

I will then use a toy numerical model (simple graphite), to see how
these concepts are realized in an actual physical system, and
a very simple approximaton to doping (rigid-band)
to simulate the effect of physical doping and detect the sources
of \ep\ interaction in graphite-like materials.
Both models are useful for a first exploration,
but inadequate to make accurate predictions for actual superconductors,
where the doping is usually obtained via chemical substitution,
which causes a major rearrangement of phononic and electronic
states, and hence sizable changes in the
values of the \ep\ interaction and \tc.

For simple
graphite, nature provides two simple realizations
of two of its sources of dormant \ep\ interactions:
\mg, a prototype of {\em covalent} superconductors, and graphite intercalated compounds, where superconductivity is correlated with the filling of {\em interlayer} states.
In Sect.\ref{sect:mat:mgb2} and \ref{sect:mat:GIC} I will discuss these
two examples in detail, and also indicate the main theoretical predictions and
experimental discoveries which were inspired by them.

In particular, the line of research
on covalent superconductors culminated in the discovery of
high-\tc\ conventional superconductivity at extreme
pressures in \sh, discussed in Sect.~\ref{sect:mat:hydrides},
which also represents a fundamental step forward in the direction
of the search of new superconductors using first-principles methods.

After describing the incredible evolution of
the state of research in conventional superconductors,
in Sect.~\ref{sect:mat:FESC} I will present a representative
example of unconventional superconductors, iron pnictides and chalcogenides (Fe-based superconductors),  discovered in 2008, which shares
many similarities with the high-\tc\ cuprates.
This example will allow me to give an idea of the many challenges that theory
faces in the description of unconventional superconductors, already
in the normal state, and currently represent a fundamental obstacle to
the derivation of numerical methods to compute \tc's.
\subsection{Conventional Superconductors: Search Strategies}
\label{sect:mat:conventional}
\begin{svgraybox}
  {\bf Matthias' rules}
  The so-called Matthias' rules are a set of empirical rules 
  that summarize the understanding of superconductors
  in the 70's.
  The rules were allegedely formulated, and revised several times, by Bernd Matthias, one of the leading material scientists in superconductivity: they are usually cited in this form:
 \begin{enumerate}
\item High symmetry is good, cubic symmetry is the best.
\item High density of electronic states is good.
\item Stay away from oxygen.
\item Stay away from magnetism.
\item Stay away from insulators.
\item Stay away from theorists.
\end{enumerate}
\end{svgraybox}

Matthias' rules 
were inspired by A15 superconductors (cubic transition metal alloys,
which can be easily doped, exhibit sharp peaks in the electronic DOS,
and are prone to lattice and magnetic instabilities),
and clearly disproved by subsequent experimental
discoveries: cuprates and pnictides (first, third, fourth and fifth rule),~\cite{SC:Bednorz_ZPB_1986,fesc:kamihara_JACS_2008}
but also conventional superconductors
such as  \mg~\cite{SC:akimitsu_mgb2}
(first and second rule).~\cite{H:DrozdovEremets_Nature2015}
However, their impact on superconductivity research has been so important that
they are still sometimes cited as arguments against conventional superconductivity or the possibility of theoretically predicting new superconductors,
together with another old quasi-empirical rule, the Cohen-Anderson limit.

In order to derive more general, non-empirical strategies to search
for new superconductors, I will begin with a simple analytical model. 
Instead the full complexity of the electronic and vibrational
properties of real superconductors, for the moment I consider an
ideal case,
in which a single phonon branch with frequency
$\omega$ and a single electronic band with Density of States at the Fermi level $N(E_F)$ are coupled through an average matrix element $I$.
In this case, the T$_c$  is well described by the Mc-Millan-Allen-Dynes formula (Eq.~\ref{eq:McMillan}) with all constant factors
set to one, $\omega_{\ln}=\omega$, and the coupling constant is
given by the Hopfield expression:  $\lambda=(N(E_F) I^2)/(M\omega^2 )$ :
\begin{equation}
  T_c=\frac{\omega}{k_B}\exp\left[-\frac{(1+\lambda)}{\lambda-\mu^{*}(1+\lambda)}\right]~,
\label{eq:Tcsimple}
\end{equation}

These formulas indicate that there are three main strategies to optimise the critical temperature of a conventional superconductor: ($i$) maximize the value of the the electronic DOS at the Fermi level $N(E_F)$ (first two Matthias' rules); ($ii$) select compounds which contain light elements (and stiff bonds), to maximise the characteristic lattice energy scales ($\omega$);
and ($iii$) increase the \ep\  matrix elements ($I$).

While the first two strategies were already understood in the early 70's,
it became apparent only with the \mg\ discovery
that it is possible to find compounds where the instrinsic
\ep\ matrix elements $I$ are much larger than in transition metals and their alloys, where the maximum \tc\ does not exceed 25 K.
In a seminal paper, An and Pickett~\cite{SC:mgb2:pickett}
pointed out that the (relatively) high \tc\ of \mg\
occurs because of ``{\em covalent bonds driven metallic}''.

\begin{figure}[h!]
\includegraphics[scale=.3]{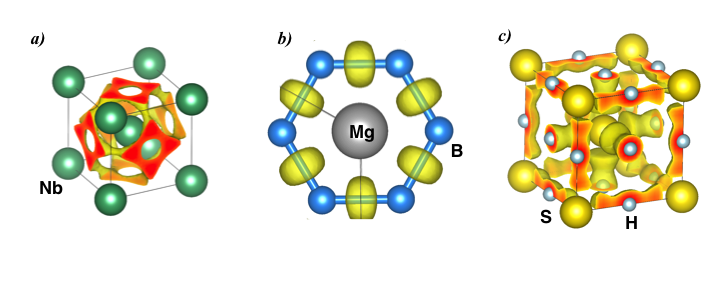}
\sidecaption[t]
\caption{Electronic localization function in conventional superconductors: bcc Nb (\tc=9 K); \mg (\tc=39 K); \sh (\tc=203 K).}
\label{fig:elf}       
\end{figure}

The reason why covalent metals have larger intrinsic \ep\
coupling than ordinary metals
can be intuitively understood looking at Fig.~\ref{fig:elf},
which shows isosurfaces of the Electronic Localization Function (ELF) for three different superconductors: Nb (\tc=9K), \mg\ (\tc=39 K) and \sh\ (\tc=203K). The ELF indicates regions where electrons are concentrated. It is clear that in \mg\ and \sh, the electrons localize along the bonds, while in Nb they are delocalized over the whole volume. When atoms undergo phonon vibrations, electrons localized along a bond will feel a much stronger perturbation than those spread out over
the whole crystal.
However, the arguments above are oversimplified, as the existence of
strong directional bonds is a necessary prerequisite for large \ep\ coupling, but it is not sufficient: due to the small energy scales involved in the
superconducting pairing, it is also essential that the electronic states which contribute to this bond lie at the Fermi level, otherwise they remain
{\em dormant}, and do not contribute to the superconducting pairing.
More precisely, this means that shifting the position of the Fermi level ($E_F$)
selects different matrix elements $g$ in Eq.~(\ref{eq:alpha_phonon})
when performing averages over the Fermi surfaces in 
in Eqs.~(\ref{eq:ME1}), (\ref{eq:ME2}), and the \ep\ interaction $\lambda$,
and hence \tc,
is appreciable only for some positions of $E_F$.

A possible strategy to search for new conventional superconductors 
thus amounts to identifying, first, possible {\bf dormant \ep\ interactions} within a given material class and, second, physical mechanisms to activate them,
such as doping, pressure, and alloying. 
{\em Ab-initio} approaches permit to explore both steps of this process,
at different levels of approximation. In the following I will illustrate
the basic working principle, using an example (simple graphite) and
two practical realizations (\mg\ and intercalated graphites);
I will also refer to the same principles when discussing possible perspective of future research in Sect.~\ref{sect:outlook}.

Before moving on with the discussion, I need to introduce a second
concept which is crucial in the search for new conventional superconductors:
{\bf lattice instabilities}.
This argument is important because it is at the heart of
the Cohen-Anderson limit.~\cite{Th:Cohen_anderson}

According to Eq.~(\ref{eq:Tcsimple}), \tc\ can apparently be increased
indefinitely, increasing $\lambda$, which contradicts what is observed in practice, since the critical temperatures of actual superconductors are
limited.
However, in my discussion I have so far disregarded
the {\em feedback} effect between phonon frequencies and \ep\ 
interaction, which is one of the main limiting factors to high-\tc\
in actual materials.
Indeed, the same \ep\ interaction
that pairs electrons leading to superconductivity
also causes a decrease (softening) of the phonon frequencies.
This means that the frequency $\omega$ appearing in Mc-Millan's formula
for \tc\ (Eq.~\ref{eq:Tcsimple}) should be more correctly rewritten as: $\omega^2=(\Omega_0)^2(1-2 \lambda_0)$, where $\Omega_0$ is the bare frequency of the lattice, in the absence of \ep\ interaction, and $\lambda_0$ is the corresponding coupling constant.
It is now easy to see that the $T_c$ for a fixed $\Omega_0$ has a maximum for a given value of $\lambda_0$ and then decreases approaching a lattice instability ($\omega \to 0$); this means that within a given material class,
\tc\ can only be increased up to a threshould value determined by
$\lambda_0$, before incurring in a lattice instability.
Using typical parameters for the A15 compounds,
which were the best known superconductors when the Cohen-Anderson
limit was formulated, gives a maximum value of \tc\ of $\sim 25$ K.
However, covalent superconductors such as \mg, doped diamond and \sh\
have much larger characteristic phonon scales ($\Omega_0,\lambda_0$),
and can substain much larger \tc.

\subsection{Dormant $ep$ interaction in graphite}

\begin{figure}[h!]
\begin{center}
  \includegraphics[width=7cm,clip,angle=0]{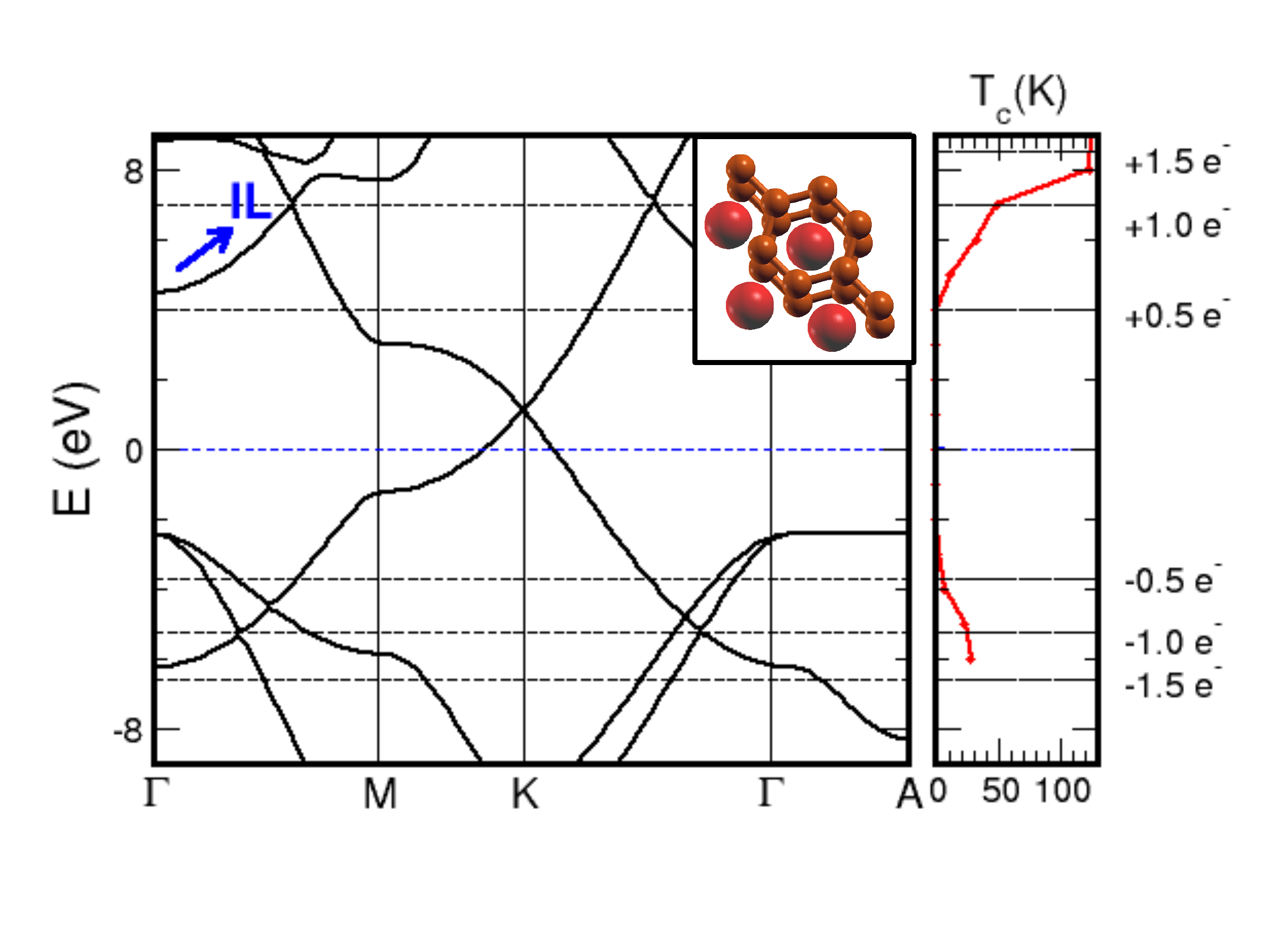}
  \caption[]{Rigid-band study of superconductivity in simple graphite; dashed
  lines indicate the position of the Fermi level for a semi-integer doping
  of holes or electrons; the blue arrow marks the position of the interlayer band, whose Wannier function is shown in the inset.
  \label{fig:graphr}
  }
  \end{center}
\end{figure}

To put general arguments on more physical grounds, we now 
consider a toy model based on an actual physical system,
simple graphite, which is realized stacking several
layers of graphene on top of each other (the stacking is thus $AAA$, in constrast to the $ABA$ and $ABC$ stacking of Bernal and rhombohedral graphite);
for this experiment, we keep the interlayer distance equal to that
of Bernal graphite.

The unit cell contains two inequivalent carbon atoms, that form a
hexagonal lattice. This means that the four  orbitals
of carbon will split into three $sp^{2}$ hybrids and a $p_z$ orbital,
forming six $\sigma$ and two $\pi$ bands, of bonding and anti-bonding
character.
The blue arrow points to a fifth band, which has no carbon character: this is the so-called interlayer (IL) band, 
which is essentially a free-electron state confined between the carbon layers.
The relative Wannier function, shown in the inset of the figure,
is indeed centered in
the middle of the interstitial region between the graphitic layers,
and has spherical symmetry.~\cite{mine:Boeri_PRB_2007,SC:csanyi_nphys_2005}

The phonon spectrum 
of pure graphite (not shown) is even simpler than its electronic
structure: In-plane bond stretching optical phonons form a rather narrow
band between 150 and 200 meV, while optical out-of-plane and acoustical modes
are more dispersive and extend up to 100 meV.

For the doping of pure graphite, corresponding to four electrons/carbon,
the Fermi level cuts the band structure near the crossing point between $\pi$ and $\pi^*$ bands, the \ep\ coupling is extremely low, and gives
rise to a negligible \tc.
However, different {\em dormant} \ep\ interactions can be activated if,
with a {\em gedankenexperiment}, the Fermi level is shifted to higher or lower
energies. This rigid band shift is the simplest approximation to physical
doping within an {\em ab-initio} calculation.

The right panel of Fig.~\ref{fig:graphr} shows  a rigid-band
calculation of the \tc\ of simple graphite, in an energy range
of $\pm$ 8 eV around the Fermi level,
corresponding to a doping of $\pm$ 1.5 electrons/carbon;
in this calculation,
the phonons and \ep\ matrix elements are computed only once, for the
physical doping of four electrons/carbon, but the averages and sums on the Fermi
surface in Eqs.~(\ref{eq:ME1}-\ref{eq:ME2}) are recomputed for different
positions of the Fermi level ($E_F$).

The figure clearly shows that \tc\ is still negligible for small variations of energy around the original Fermi level,
but has a marked increase as soon as holes or electrons are doped into
the $\sigma$ (bonding or antibonding) or IL bands, reaching a maximum
of $\sim 100$ K, when the Fermi level reaches a large van Hove
singularity at $\sim 8$ eV, corresponding to the bottom of the $\sigma^*$
band. 
Apart from causing substantial deviations in \tc's, the coupling of phonons to
$\sigma$, IL and $\pi$ electrons is also
qualitatively different.
$\sigma$ and $\pi$ electrons couple mostly to high-energy optical
phonons, which modulate the interatomic distances in the hexagonal
layers, and hence the hopping between neighboring carbon sites;
IL electrons, instead, respond to out-of-plane phonons, which modulate
their overlap with the $\pi$ wavefunctions that stick out-of-plane.
This effect causes a substantial variation in the shape of the
Eliashberg functions (not shown) and $\omega_{log}$.
The right scale on Fig.~\ref{fig:graphr} indicates the position of the
Fermi level for semi-integer values of hole(-) or electron(+) doping;
in both cases, a finite \tc\ is obtained for 
a minimum doping of half an  electron/carbon is needed,
but obtaining a high-\tc\ requires a much higher doping ($\sim$ 1 $e^{-}$/C atom). Note that the highest doping levels that can be obtained with field effect using liquid electrolytes are much lower, i.e. of the order of a few tenths of $e^{-}$/carbon, so that the only viable alternative to realize superconductivity
in doped graphite is via chemical doping.

The rigid-band approximation is instructive to identify dormant
\ep\ interactions, but the calculated \tc's are often severely
overestimated compared to other experiments or more
sophisticated approximations for doping.
In fact, the rigid-band approach neglects important effects due
to the self-consistent rearrangement of electrons produced by doping,
such as band shifts, renormalization of the phonon frequencies, and screening
of the \ep\ matrix elements.~\cite{mine:Subedi_PRB_2011,SC:Casula_picene_PRB_2012}

For simple graphite, nature provided two extremely ingenious realizations of our predictions, discussed in the next two sections.

\subsection{Magnesium Diboride and other Covalent Superconductors:}
\label{sect:mat:mgb2}
\begin{figure}[h!]
\includegraphics[width=0.6\textwidth,clip,angle=0]{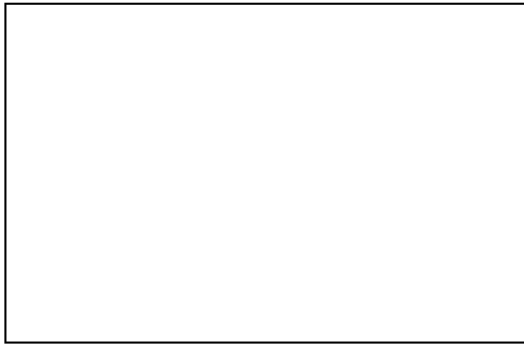}
  \sidecaption[t]
\caption[]{Two-gap superconductivity in \mg: Anisotropic distribution of the gap on the Fermi surface, predicted by DFT ($a$)~\cite{SC:mgb2_choi_2002};
  Evidence of two-gap behaviour from tunneling experiments.($b$)~\cite{SC:mgb2_gonnelli_PRL2002}.
  \label{fig:mgb2}}
\end{figure}

With a \tc\ of 39 K, magnesium diboride (\mg) currently holds the record
for conventional superconductivity at ambient pressure.~\cite{SC:akimitsu_mgb2,SC:qcarbon_2017}
Its crystal structure is layered: boron forms graphite-like hexagonal
planes; magnesium is placed in-between, at the center of the hexagons.
Mg is completely ionised (Mg$^{2+}$), and thus \mg\ is not only isostructural,
but also isoelectronic to graphite.
However, due to the attractive potential of the Mg ions, the center
of mass of the $\pi$ bands is shifted up with respect to that of the $\sigma$
bands compared to simple graphite,
so that \mg\ behaves effectively as a compensated (semi)-metal.
The $\sigma$ holes and $\pi$ electrons
form two cylinders around the center 
and a 3D tubular network around the corners of the Brillouin zone, respectively,
as shown in Fig.~\ref{fig:mgb2}
($a$).

The two group of electronic states
experience a rather different coupling to phonons:
over 70$\%$ of the total \ep\ coupling,
in fact, comes from $\sigma$ holes and
bond-stretching phonons; the rest is distributed
over the remaining phonon modes and electronic states.~\cite{SC:mgb2:kong}
The simplest approximation to account for the strong anisotropy of the \ep\
coupling over the Fermi surface is to replace the \ep\ coupling constant
$\lambda$ in Eq.~(\ref{eq:ME1}-\ref{eq:ME2}),
with $2\times2$ matrices of the form:~\cite{SC:mgb2_clean_or_dirty}
\begin{equation}
  \lambda=\left(\begin{array}{cc}
    \lambda_{\sigma \sigma} & \lambda_{\sigma \pi}\\
    \lambda_{\pi \sigma} & \lambda_{\pi \pi}\\
    \end{array}
  \right)=
  \left(\begin{array}{cc}
    1.02 & 0.30\\
    0.15 & 0.45\\
    \end{array}
  \right)
\label{eq:lambda_2gap}
\end{equation} 
When the interband coupling is finite but appreciably smaller than the intraband
one, $\left|\lambda_{\sigma \pi}+\lambda_{\pi \sigma}\right| < \left|\lambda_{\sigma \sigma}+\lambda_{\pi \pi}\right|$, the theory predicts that
  experiments should observe two distinct gaps,
  closing at the same \tc.~\cite{Th:Suhl_Matthias_PRL_1959}
  Two gap superconductivity was indeed observed by different experimental techniques: specific heat, tunneling, Angle-Resolved Photoemission Spectroscopy (ARPES) for the first time in \mg.~\cite{SC:mgb2_clean_or_dirty}
  
  Indeed, in most superconductors, multiband and anisotropic effects are
  extremely difficult to detect because they are suppressed by
  the interband scattering induced
  by sample impurities; in \mg\ 
  the real-space orthogonality of the $\sigma$ and $\pi$
  electronic wavefunctions prevents interband scattering,
  and two-gap superconductivity can be detected also by techniques with a limited resolution.~\cite{SC:mgb2_clean_or_dirty}

  Fig.~\ref{fig:mgb2}($b$) shows in- and out-of-plane
  tunneling spectra of \mg, which permit to unambiguously identify
  a $\sigma$ (large) and $\pi$ (small) gap.~\cite{SC:mgb2_gonnelli_PRL2002}
  The experimental spectra compare extremely well with the theoretical prediction of two different gaps on the $\sigma$ and $\pi$
  sheets of the Fermi surface; the image in the left panel of the
  figure shows an anisotropic DFT-ME calculation of the gap; in the figure, the color of the Fermi surface is proportional to the size of the gap on that sheet.~\cite{SC:mgb2_choi_2002}

%

  Besides providing the first clear case of two-gap superconductivity,
  \mg\ is the first example of superconductivity from {\em doping covalent
    bonds}.
The most spectacular realization of this possibility  came in 2004,
with the report of a \tc\ of 4 K in heavily boron-doped diamond,
raised to 11 K in thin films.~\cite{SC:diamond:Ekimov_2004,SC:Takano_diamond_APL_2004}

Diamond is a wide-band insulator ($\Delta \sim 5.5$ eV);
when boron is doped at small concentrations, an acceptor band forms within the gap. 
At the much higher doping levels ($\sim 1-10\%$) realized
in superconducting samples, the impurity band overlaps so strongly with the valence band of diamond, that the net effect of B doping is to create
holes at the top of the valence band.~\cite{SC:diamond_Yokoya_nature_2005} 
This band is
formed by the bonding combination of the four carbon $sp^3$ hybrids.
Boron-doped diamond can thus be seen as a 3D analogue of \mg,
where, similarly to \mg,
a small fraction of $\sigma$ holes created by B doping exhibits
an extremely strong coupling to bond-stretching phonons.

Since the \ep\ coupling is concentrated in a single type
of phonon modes and electronic states, the simplified formulas
introduced in Sect.~\ref{sect:mat:conventional} give a reasonable approximation
to \tc\ for both \mg\ and
diamond.~\cite{mine:Boeri_PRL_2004,SC:Pickett_diamond_PRL_2004,SC:Blase_PRL_2004,SC:diamond_giustino_PRL_2007,SC:diamond_hoesch_PRB_2007}
Furthermore, they permit to understand why, even though the
C-C bonds in diamond are actually stiffer than the B-B bonds in \mg,
the measured \tc's are much lower.
In fact, near the bottom (or top) of a band, i.e. at low dopings, the DOS
increases as  $\sqrt{E}$ in 3D, while it is virtually constant in 2D.
This implies that, as holes or electrons are doped
into the system, $N(E_F)$, and hence $\lambda$, increases much more
slowly in 3D systems than in 2D ones. Thus, for
physical ranges of dopings the maximum \tc\ in doped semiconductors
is typically very low.~\cite{mine:Boeri_PRL_2004,SC:silicon:Bustarret_2006}

  \begin{figure}[h!]
\begin{center}
    \includegraphics[width=6cm,clip,angle=0]{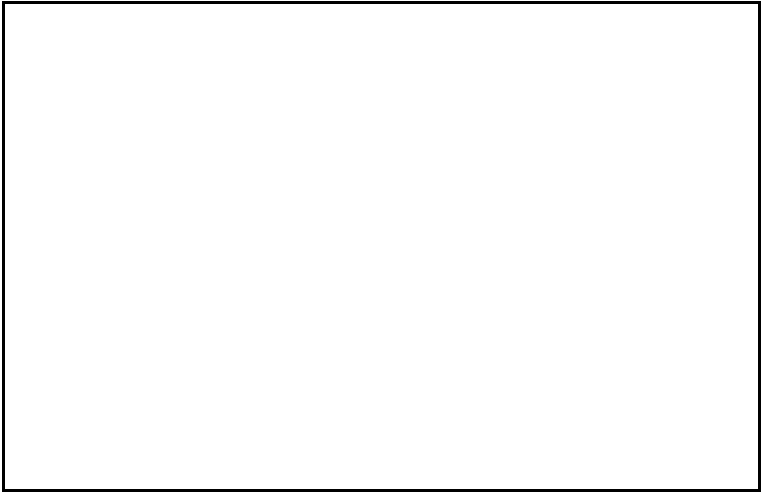}
  \caption{Electron-phonon coupling (top) and \tc\ (bottom) as a function of
    doping in different carbon-based superconductors. Adapted from
    Ref.~\cite{SC:giustino_PRL_2010}.}
  \label{fig:graphane}
  \end{center}
\end{figure}

A very ingenious strategy attain high-\tc\ conventional superconductivity in a doped $sp^3$ system avoiding negative dimensionality effects was proposed a few years later by Savini et al.~\cite{SC:giustino_PRL_2010},
i.e. realizing superconductivity in p-doped graphane (fully hydrogenated graphene). Graphane can be considered a 2D version of diamond, in which the bonding is still $sp^3$, hence the matrix
elements are as large as those of diamond, but the DOS is 2D,
hence $\lambda$ is sizable already at low dopings.
For this compound, the authors of Ref.~\cite{SC:giustino_PRL_2010}
estimate a \tc\ of $\sim 90$ K already for 1$\%$ doping. Figure~\ref{fig:graphane}, from the original reference, compares the behaviour of \tc\
with doping in the difference class of carbon-based superconductors discussed so far.

 Although not as spectacular as graphane, in general many carbon-based compounds
 doped with boron are good candidates for superconductivity with relatively
 high temperatures.
 For example, recent experiments indicate that Q-carbon,
 an amorphous form of carbon, in-between diamond and graphite,
 can achieve \tc's as high as 56 K upon boron doping.~\cite{SC:qcarbon_2017}
 
 A complementary idea is that of doping boron-rich phases with carbon; boron, being an electron-deficient
 material, forms a variety of structures with two- and three-center bonds.
 One of the most common motifs is the icosahedron ($B_{12}$), found, for example, in the $\alpha$ and $\beta$ phases of elemental boron, as well as in
 superconducting dodecaborides.~\cite{SC:Matthias_B12}
 Boron icosahedra doped with carbon are predicted to superconduct
with \tc's as high as $37$ K.~\cite{SC:mauri_B12}

\subsection{Intercalated Graphites}
\label{sect:mat:GIC}

\begin{figure}[h!]
\includegraphics[width=12cm,clip,angle=0]{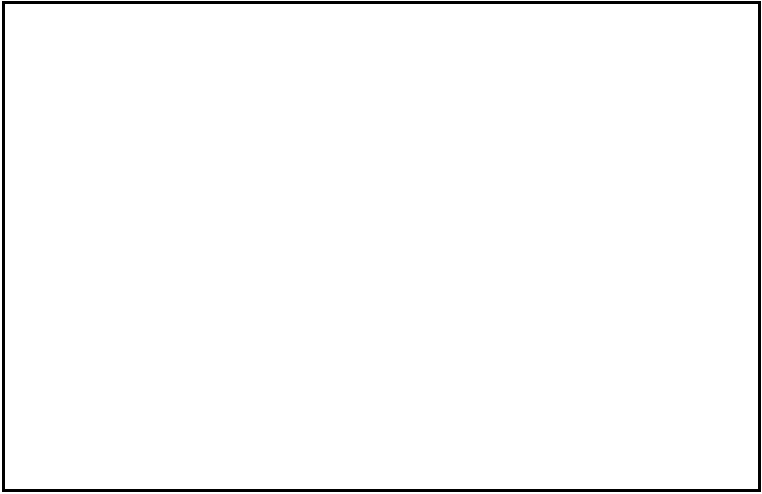}
\caption[]{Physical properties of superconducting graphite intercalation compounds:
  ($a$) Dependence of \tc\ on the interlayer separation in different GICs;~\cite{mine:kim_PRL_2007};
  ($b$)Anisotropic superconducting gap predicted from SCDFT calculations in CaC$_6$,~\cite{SC:Sanna_PRB_2007};
    ($c$) Superconducting gap measured by ARPES in Li-decorated
  graphene.~\cite{SC:Ludbrook_Ligraphene_PNAS2015}
  \label{fig:GIC}}
\end{figure}

While \mg\ can be considered a natural realization of hole-doped
graphite $\sigma$ bonds, superconducting graphite intercalation compounds
(GICs) are the practical realization of superconductivity by doping
into the interlayer states. Although the \tc's are substantially
lower than in covalent metals, this type of superconductivity is
quite interesting, because it can be more easily manipulated
by external means (doping, pressure), and also
realized in ``artificial'' systems, such 
mono- and bi-layers of graphene decorated with alkali or noble metals.

Intercalated graphites had been extensively studied in the 70’s, because of their high mobilities, but the only known superconducting member
of the family, KC$_8$, had an extremely small \tc ($< 1 K$).
Only in 2005, superconductivity with relatively high \tc's
was reported in several $A$C$_6$ compounds: CaC$_6$ (\tc=11.5 K), YbC$_6$
(\tc=6.5 K), and later SrC$_6$ (\tc=1.6 K).~\cite{SC:weller_nphys_2005,mine:kim_PRL_2007}
At the same time, it was observed that for all newly-discovered
members of the family, 
the occurrence of superconductivity clearly correlates with the filling
of the interlayer band, which is empty in non-superconductors.~\cite{SC:csanyi_nphys_2005}

A strong correlation is also found between the \tc\ and the distance
between two subsequent graphitic layers in different $A$C$_6$ compounds
-- Fig.~\ref{fig:GIC}$(a)$.
This can be easily explained within the
conventional \ep\ scenario,~\cite{SC:calandra_PRL2005,mine:kim_PRB_2006,mine:Boeri_PRB_2007,mine:kim_PRL_2007}
since, as demonstrated in Sect.~\ref{sect:mat:conventional}, when localized
due to doping and confinement effects, interlayer electrons can couple
to $\pi$ electrons through out-of-plane phonons of the carbon layers
(in GICs, additional coupling is also provided by intercalant modes).

Like in \mg, the distribution of the \ep\ interaction, and hence of the superconducting gap, is very anisotropic on the Fermi surface, being much larger
for the interlayer electrons  -- central sphere in
Fig.~\ref{fig:GIC}($b$) --  than for  $\pi$ ones, which form the outer
tubular manifold.

The idea superconductivity due to the localization of interlayer states was exploited in Ref.~\cite{SC:Profeta_Graphene_Nat2012}, proposing to achieve to superconductivity  in Li-decorated graphene. A \tc\ of 5.9 K, close
to the theoretical prediction of 8.6 K, in 2015 in Ref.~\cite{SC:Ludbrook_Ligraphene_PNAS2015}. Fig.~\ref{fig:GIC}($c$) shows the superconducting gap, measured by ARPES. Note that also in this case there is a visible
variation of the gap between $\pi$ and IL sheets of the Fermi surface.

\subsection{High-\tc\ Conventional
  Superconductivity in High Pressure Hydrides}
\label{sect:mat:hydrides}

The most spectacular realization of high-\tc\
conventional superconductivity from doped covalent bonds
can be found in \sh,
which, so far, holds the \tc\
the record among all (conventional and unconventional) superconductors.

Indeed, superconductivity at high pressures is a rather ubiquitous phenomenon, because high pressures tend to increase the hopping between neigbouring sites and
hence metallicity.
Almost all the elements of the periodic table can be made superconducting; the typical \tc's are rather low, reaching a maximum of $\sim$ 20 K in Li, Ca, Sc, Y, V, B, P, S, ~\cite{SC:HAMLIN_highp_physicaC_2015} well reproduced by {\em ab-initio}
  calculations.~\cite{SC:Profeta_Li_PRL2006,SC:Martonak_Ca_PRL2009,SC:Monni_S_PRB2017,mine:Flores_PRM2017}

  Hydrogen and its compounds represent a notable exception, having provided, in 2014, the first example of high-\tc\ conventional
  superconductivity. This discovery
   was the coronation of a 50-years long search,
  inspired by two insightful papers by Neil Ashcroft, predicting high-\tc\
  superconductivity in metallic hydrogen (1968) and covalent
  hydrides (2004).~\cite{H:Ashcroft_PRL1968,H:Ashcroft_PRL2004}
Both predictions rely on the general arguments for high-\tc\ conventional superconductivity introduced in section ~\ref{sect:mat:conventional}, i.e.
large phonon frequencies due to the light hydrogen mass ($ii$)
combined with large
\ep\ matrix elements due to the lack of screening from core electrons ($iii$)
can yield remarkable superconducting transition temperatures, even if the Density of States at the Fermi level is moderate ($i$).

In 1968, the first of Ashcroft's predictions,
superconductivity in metallic hydrogen, seemed merely an academic speculation.
The pressure required to metallize hydrogen, which is a large gap insulator at ambient pressure ($\Delta \sim 10$ eV),
was clearly beyond reach for the experimental techniques of the time.
However, fifty years later, high-pressure research has advanced to such a point
that at least three groups have reported evidences of hydrogen metallization, at pressures ranging from 360 to 500 GPa (3.6 to 5 Mbar).\cite{H:Dalladay_H_Nature2016,H:Eremets_H_arxiv2016,H:Dias_science2017}

These reports are still controversial, since the pressure ranges are close to
the limit of current high-pressure
techniques.~\cite{H:eremets_comment,H:goncharov2017comment} 
The insulator-to-metal transition has two possible origins:  band overlap,
in the molecular $CmCa$ phase, or a structural transition
to an atomic $\beta-Sn$ phase. The two phases are almost impossible to
discern experimentally because hydrogen is a poor X-Ray scatterer, and
theoretical studies also predict an unusual spread of values (300-500) GPa
for the transition boundary between the two phases,
due to the different approximations used to account for quantum lattice
effects.~\cite{H:McMahon_RMP_2012}
However, according to DFT calculations, both the $CmCa$ and the $\beta-Sn$ structures should become superconducting at or above ambient temperature, suggesting that the first of Ashcroft's predictions may soon be realized.~\cite{H:Cudazzo_hydrogen_2008,H:McMahon_PRB_2011,H:Borinaga_PRB_2016}

The second prediction, superconductivity in covalent hydrides, was experimentally verified
at the end of 2014.  The underlying idea is that in covalent hydrides, metallization of the hydrogen sublattice should occur at lower pressure than in pure
hydrogen, because the other atoms exert an additional chemical pressure.

Indeed, in 2008, superconductivity was measured
in compressed silane (SiH$_4$) at $120$ GPa, but the measured \tc (17 K) was disappointingly low compared to theoretical prediction of
100 K.~\cite{H:silane_exp_2008,H:Li_silane_PNAS2010}
However, this finding proved that  the experimental knowledge to make covalent hydrides was available, and it was only a matter of time before high-\tc\ conventional superconductivity would actually be observed.
{\em Ab-initio} calculations revealed two essential missing
pieces of the puzzle:
$a$) Megabar pressures can stabilize {\em superhydrides}, i.e.
phases with much larger hydrogen content than the hydrides stable at ambient pressure; $b$) some of these {\em superhydrides} are metals with unusually strong bonds, which can lead to high-\tc\ superconductivity.~\cite{H:zurek_PNAS_2009}

The first high-\tc\ superconductor discovered experimentally is a sulfur superhydride (\sh), which is stabilized under pressure, by compressing
gaseous sulfur dihydride (SH$_2$) in a hydrogen-rich atmosphere. The compound metallizes at $\sim 100$ GPa, where it exhibits
a \tc\ of 40 K. The \tc\
increases reaching a maximum of over 203 K at 200 GPa;
isotope effect measurements confirmed that superconductivity is of conventional origin.
I refer the reader to the original experimental~\cite{H:DrozdovEremets_Nature2015,H:Goncharov_SH3_exp_2016,H:Einaga_SH3_nphys_2016,H:Capitani_SH3_nphys_2017}
and theoretical~\cite{H:Duan_SciRep2014,mine:Heil_PRB_2015,H:FloresSanna_H3Se_EJPB2016,H:errea_PRL_2015,H:errea_nature_2016,H:Bernstein_PRB_2015,H:Quan_PRB_2016,H:Akashi_magneli_PRL_2016,H:Sano_SH3_PRB_2016} references for a more detailed discussion of specific aspects of the \sh\ discovery, such as the nature and thermodynamics of the SH$_2$ to \sh\ transition, anharmonic and non-adiabatic effects.~\cite{H:Gorkov_SH3_RMP_2018}

In the history of superconducting materials, \sh\ stands out for one main reason:
it is the first example of a high-\tc\ superconductor
whose chemical composition, crystal structure and superconducting transition
temperature were predicted from first-principles before the actual
experimental discovery. In fact, a few months before the experimental report,
Duan et al.~\cite{H:Duan_SciRep2014} predicted
that SH$_2$ and H$_2$ would react under pressure and give rise to a highly symmetric bcc structure, later confirmed by X-Ray experiments,~\cite{H:Einaga_SH3_nphys_2016} which could reach a \tc\ as high as 200 K at 200 GPa.

From the point of view of electronic structure,
\sh\ is a paradigmatic example of high-\tc\
conventional superconductivity.
Indeed, in this case all three conditions reported in Sect.~\ref{sect:mat:conventional} are verified:
\sh\ is a hydride, and hence its characteristic vibrational frequencies are high ($i$); the Fermi level falls in the vicinity of a van Hove singularity of the DOS of the bcc lattice ($ii$); and the \ep\ matrix elements are high due to the unusual H-S
    covalent bonds stabilized by pressure ($iii$).

   Replicating a similar combination is not simple, even in other high-pressure hydrides, where one can hope that high pressure may help stabilize unusual bonding environments and hydrogen-rich
stoichiometries.
   Despite an intense theoretical exploration of the high-pressure
   superconducting phase diagrams of binary hydrides,~\cite{H:zurek_PNAS_2009} only a few candidates
   match or surpass the \tc\ of \sh: Ca, Y, La.~\cite{H:Wang_PNAS2012_CaH6,H:Liu_PNAS_la_hydrides}

   In fact, the formation of covalent directional bonds between hydrogen and other atoms appears to be very sensitive to their electronegativity difference,~\cite{H:Bernstein_PRB_2015,H:Fu_Ma_pnictogenH_2016} and in binary hydrides the possibilities to optimize this parameter are obviously limited.
A possible strategy to overcome this limitation is to explore ternary hydrides, where the electronegativity, atomic size, etc. can be tuned continuously combining different elements. However, since ternary Gibb's diagrams are computationally much more expensive than binary convex hulls, fully {\em ab-initio} studies of ternaries are rare.
Two recent studies explore two different strategies towards high-\tc\ in ternary hydrides: doping low-pressure molecular phases of covalent binary hydrides, like water;~\cite{H:Flores_H2O}
and off-stoichiometry phases of alkali-metal alanates and borates,
which permit to independently tune the degree of metallicity
and covalency.~\cite{mine:Kokail_PRM_2017}

Due to the intrinsic difficulty of reaching Megabar pressures, and the
limited information that can be extracted from X-Ray spectra,
the experimental information on high-pressure hydrides is much
scarcer than theoretical predictions.
Nevertheless, there has been a substantial progress in the last five years: Metallic superhydrides predicted by theory have been reported
in hydrides of alkali metals (Li,Na),~\cite{H:Pepin_PNAS_Li_2015,H:Struzhkin_Na_Natcomm_2016} transition metals (Fe),~\cite{H:Pepin_FEH_2017} group-IV elements, etc. For some of these systems, first-principles calculations predict
substantial superconducting temperatures,~\cite{H:zurek_PNAS_2009} while other cases are controversial.~\cite{H:Majumdar_FEH,H:Oganov_FEH,H:Heil_FEH_2018}

Resistivity and susceptibility measurements required to detect superconductivity under pressure are even more challenging, and therefore the available information on superconductivity in high-pressure hydrides is still very scarce:
besides \sh, only one other hydride, \ph, has been shown to superconduct at high pressures, albeit with a lower \tc\ ($\sim 100$ K).~\cite{H:Drozdov_PH3_arxiv2015}
At variance with \sh, \ph\ is highly metastable, and samples rapidly degrade over time, consistently with {\em ab-initio} predictions of metastability.~\cite{H:Fu_Ma_pnictogenH_2016,mine:Flores_PRB_2016,H:shamp_decomposition_2015}
SeH$_3$, which should exhibit \tc's comparable to \sh, has been succesfully synthesized at the end of last year~\cite{H:Pace_selenium_JPCP_2017}, but superconductivity has not been measured yet. 

The same arguments that motivated the search for high-\tc\ conventional superconductivity in high-pressure hydrides can be applied to other compounds that contain light elements, such as Li, B, C etc. Some of these elements form covalent
structures with strong directional bonds already at ambient pressure, and high pressures could be used to optimize doping, stoichiometry, etc. However, element-specific factors can unpredictably affect properties relevant in high-pressure superconductivity.
For example, elemental phases of alkali
metals at high pressure exhibit a characteristic
{\em interstitial charge localization}, due to avoided core
overlap. The same effect, which is extremely detrimental for conventional
superconductivity (charge localized in the interstitial regions has an intrinsically
low coupling to phonons), also occurs in many of their compounds.
This was shown, for example, for the  Li-S system, whose behaviour at high
pressure is remarkably different from H-S.~\cite{mine:Kokail_PRB_2016}.

\subsection{Unconventional Superconductivity in Fe pnictides and chalcogenides}
\label{sect:mat:FESC}

After illustrating the remarkable progress in the research on conventional
superconductors, I have chosen to discuss 
the single largest class of unconventional superconductors,
formed by Fe pnictides and chalcogenides (Fe-based superconductors, FeSC).
The \tc's of these compounds, discovered in 2008,
go up to 56 K in the bulk, and allegedely up to 100 K in
monolayers of FeSe grown on SrTiO$_3$.~\cite{fesc:kamihara_JACS_2008,Ge_FeSeML_NatMat_2014,fesc:Paglione_review_natphys2010,fesc:johnston_AdP_2010,fesc:huang_ARCMP_2017}
Their rich phase diagram,
the quasi-2D crystal structure and the unconventional
behavior of the superconducting gaps, are strongly reminiscent
of the high-\tc\ cuprates.
Thus, I will use FeSC as a representative example to illustrate
the challenges faced by {\em ab-initio} methods in unconventional superconductors, both in the normal and in the superconducting state.~\cite{fesc:Basov_manifesto_nphys_2011}

\begin{figure}[h]
\begin{center}
  \includegraphics[width=7.5cm,clip,angle=0]{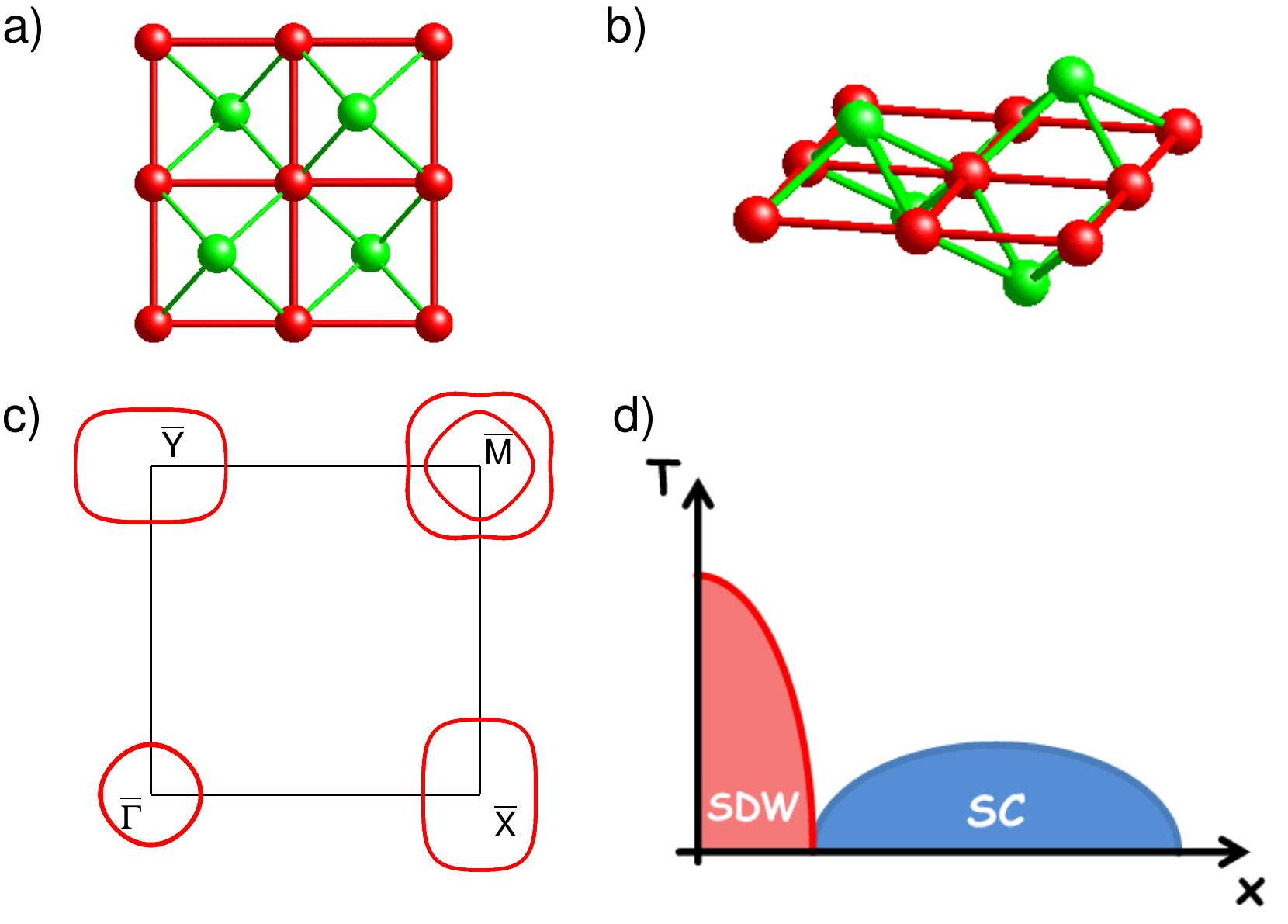}
\sidecaption[t]
\caption[]{\small{ Common features of Fe-based superconductors (FeSC): Fe$X$ layers,
seen from the top ($a$) and side ($b$); Fe and $X$ atoms are red and green, respectively.
($c$) Two-dimensional Fermi surface of LaOFeAs, from Ref.
~\cite{mine:Andersen_adp2011},
with the hole  and electron pockets centered at $\bar{X}$,
$\bar{Y}$ and $\bar{M}$ respectively. 
Here a third hole pocket at $\bar{\Gamma}$ is also present.
($d$) Phase diagram, showing the transition from SDW
to superconducting (SC) state;
 $x$ is an external tuning parameter (doping, pressure).}}
\label{fig:fesc}
\end{center}
\end{figure}

Fig.~\ref{fig:fesc} shows the main  features, common to most compounds:
\begin{itemize}
  \item
    ($a$)-($b$) A common structural motif, consisting of square planes
    of iron atoms, and $X_4$ tetrahedra;
$X$ is either a pnictogen (As,P) or a chalchogen (Se,Te).
\item
($c$) A quasi-two-dimensional Fermi surface, comprising two hole and two electron sheets, strongly nested with one other.
\item
($d$) A phase diagram, in which superconductivity sets in after suppressing a spin density wave (SDW) ordered state, by means of doping or pressure.
The most common SDW pattern is a stripe one, in which the Fe spins are aligned ferromagnetically along one of the edges of the Fe squares, and antiferromagnetically along the other.
The SDW transition is usually preceded by a structural-nematic transition.~\cite{fesc:Fernandes_nphys2014}
\end{itemize}
FeSC can be grouped in different {\em families},
depending on the nature of the Fe$X$ layers and of the intercalating blocks; the most common are: $11$ Fe chalcogenides (FeS, FeSe, FeTe); $111$ alkali-metal pnictides (LiFeAs, LiFeP, NaFeAs, NaFeP, etc); 122 pnictides (Ba/KFe$_2$As$_2$, BaFe$_2$P$_2$, CaFeAs$_2$, EuFe$_2$As$_2$ etc.); $1111$ pnictides (LaOFeAs, LaOFeP, etc); 122 chalcogenides (KFe$_2$Se$_2$, etc.). In most of these cases, superconductivity appears around a a Fe $d^6$ configuration, but it survives up to high hole or electron dopings in 122 K pnictides and chalcogenides.

In most FeSC, the superconducting gap exhibits a feature which is
distinctive of unconventional superconductors, i.e. changes sign over
the Fermi surface.
According to Eq.~\ref{eq:BCSgap}, this is only possible if, unlike
\ep\ interaction, the pairing interaction $V_{\mathbf{k},\mathbf{k}'}$ is repulsive over some regions of reciprocal space.

In contrast to the cuprates, where
the Fermi surface topology favors $d$-wave superconductivity, the gap
of most FeSC exhibits a characteristic $s^{\pm}$ symmetry, with
opposite signs on the hole and electron sheets of the Fermi surface.
However, substantial variations
in the symmetry and magnitude of the superconducting gap are observed among and within different FeSC families, which can be related to changes in the general shape and orbital distribution of Fermi surface sheets.~\cite{fesc:hirschfeld_gap_RPP_2011}

The general topology of the Fermi surface is well described by DFT
calculations; however, quasi-particle bands measured by ARPES
exhibit renormalizations and shifts, which hint to strong local
correlation effects beyond DFT, which require the use of specialized methods,
such as DFT+DMFT.~\cite{DFT:Kotliar_DFTDMFT_RMP2006}

\begin{figure}[h]
\begin{center}
  \includegraphics[width=12cm,clip,angle=0]{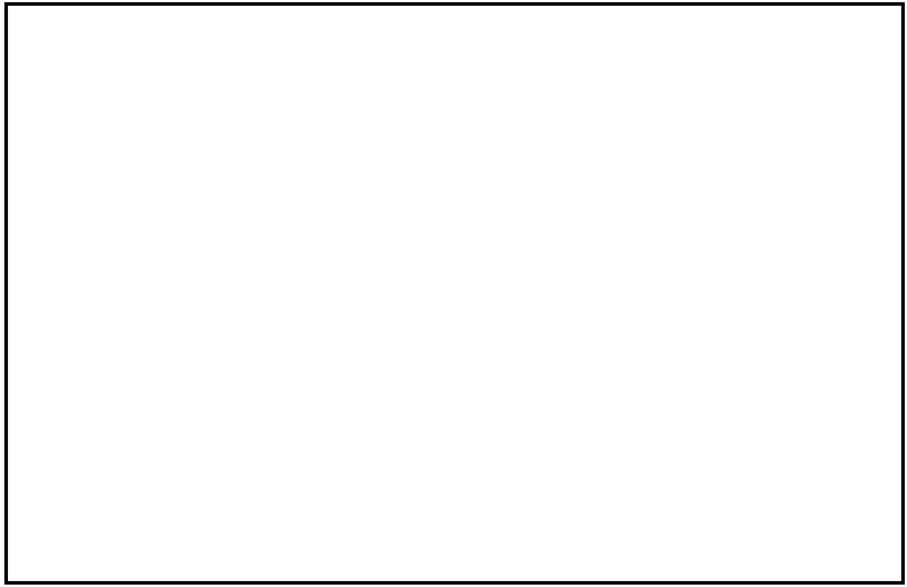}
\caption[]{Normal and superconducting-state properties of FeSC, predicted by DFT+many-body methods. $a$) Renomalization of the Fermi surfaces and ($b$) variation of the ordered and local magnetic moment across different families
of FeSC, predicted by in DFT+DMFT;~\cite{fesc:yin_nphys_2011} $c$) Superconducting trends in $1111$ pnictides predicted by DFT+RPA, Ref.~\cite{fesc:kuroki_PRB_2009}; $d$) Superconducting gap of FeSe predicted by SCDFT.~\cite{DFT:Essenberger_PRB_2016}
}
\label{fig:fesc2}
\end{center}
\end{figure}

Figure \ref{fig:fesc2} ($a$), from Ref.~\cite{fesc:yin_nphys_2011},
shows the Fermi surfaces calculated within DFT (upper panel)
and DFT+DMFT (lower panel) for representative FeSC:
LaFePO, BaFe$_2$As$_2$, LiFeAs, KFe$_2$As$_2$.
The first three are $d^6$ pnictides, with the typical hole-electron
topology shown in Fig.~\ref{fig:fesc},
eventually deformed due to interlayer hopping.~\cite{mine:Andersen_adp2011}
In KFe$_2$As$_2$, where the electron count is $d^{5.5}$, the hole pocket
is expanded and the electron pocket has lost its typical circular shape.

The figure shows that, in most cases, the inclusion of electronic
correlation modifies the Fermi surface quantitatively, but not
qualitatively, with respect to the DFT prediction.
The renormalizaton of the quasi-particle band dispersion depends
on their orbital character, being stronger for Fe  $t_2$ ($xz,yz,xy$)
than for $e$  ($x^2-y^2$, $3z^2-1$) orbitals, and on the nature of the $X$ atom; correlation effects are in fact weakest for $X$=$P$, and increase
moving to $As$,$S$,$Se$ and
$Te$.~\cite{fesc:demedici_Mott_PRL_2014,fesc:yin_nphys_2011}
Orbital-selective mass renormalizations and poor Fermi liquid behavior
are characteristic of a special correlation regime,
known as the {\em Hund's metal} regime, which has been
extensively studied by several authors;~\cite{fesc:werner_PRL2008,fesc:haule_njp2009,fesc:demedici_janus_PRL2011,fesc:yin_nphys_2011} excellent reviews can be found in Refs.~\cite{fesc:georges_ARCM_2013,fesc:demedici_hund_CM2017}.

A very important consequence of the interplay between metallicity, strong
correlations, and multiband character
in FeSC is their anomalous magnetic behavior.
Early DFT studies pointed out that magnetism cannot
be consistently described within neither the itinerant (Slater), nor the localized (Heisenberg) scenario.~\cite{mine:Mazin_PRB_2008,fesc:Johannes_magnetism_nphys_2008}
A regime which is intermediate between the two cannot
be treated in DFT, which is a mean-field theory,
but requires methods able to capture the {\em dynamics} of the
magnetic moments at different frequency scales.~\cite{fesc:yin_nphys_2011,fesc:aichhorn_PRB_2011,fesc:Hansmann_PRL_2010,fesc:toschi_PRB2011,fesc:skornyakov_PRL2011,mine:Schickling_PRL_2012}.

Panel ($c$) shows a DFT+DMFT calculation of the magnetically ordered
state for a variety of FeSC.~\cite{fesc:yin_nphys_2011}
Circles (and stars) indicate the theoretical (experimental)
value of the ordered SDW mangnetic moment,
which shows large variations among different compounds.
Chalcogenides exhibit large ordered moments, while 111 and 1111
pnictides and phosphides exhibit smaller values.
(Grey) squares indicate the value of the local magnetic moment,
which oscillates over a much smaller range of values around 2.5 $\mu_B$.
In the SDW  phase the two moments cohexist;
when the itinerant moment is suppressed with doping or pressure,
only the local moment survives, meaning that in the non-magnetic
regions of the phase diagram FeSC are in a paramagnetic state.

The superconducting state that emerges from this complicated normal
state is clearly unconventional.
According to linear response calculations, \ep\ coupling plays a
marginal role in the superconducting pairing, since the
values of the calculated coupling constant $\lambda$ in all FeSC
is extremely small, even including
magnetoelastic
effects.\footnote{This result is generally accepted, although
  DFT+DMFT studies evidenced a strong renormalization of some phonon modes,
  due to strong electronic correlations;~\cite{fesc:Mandal_epfese_PRB2014} \ep\ coupling has also been suggested to play a primary role in the enhancement of the superconducting \tc\ in FeSe monolayers grown on SrTiO$_3$,~\cite{fesc:huang_ARCMP_2017} although in this case the modes involved in the
pairing belong to the substrate.}~\cite{mine:Boeri_PRL_2008,mine:Boeri_PRB_2010,fesc:subedi_PRB2008}
The strongest candidate as superconducting mediator
are spin fluctuations, as suggested by the unconventional
symmetry of the superconducting gap and the proximity of SDW and
superconductivity in the phase diagram.~\cite{fesc:chubukov_PRB_2008,fesc:kuroki_PRL_2008,fesc:mazin_PRL2008} As discussed in Sect.~\ref{sect:theory:DFT}, there is currently no first-principles theory
of spin fluctuations with an accuracy comparable to that for \ep\ interaction.
Most studies of superconductivity in FeSC have therefore
adopted a {\em hybrid} approach,
in which the electronic structure in the vicinity of the Fermi level is downfolded or projected to an effective analytical
tight-binding model,~\cite{fesc:graser_NJP2009,fesc:kuroki_PRB_2009,fesc:miyake_JPSJ2010, mine:Andersen_adp2011}
and electron-electron interactions leading to spin and
charge fluctuations are included at a second stage 
with many-body techniques.
The most common are the Random-Phase-Approximation (RPA),
Fluctuation Exchange (FLEX), Functional renormalization Group (fRG).~\cite{fesc:chubukov_PRB_2008,fesc:thomale_PRB_2009,fesc:hirschfeld_gap_RPP_2011,fesc:platt_PRB_2012}
Although not quantitatively predictive, these {\em weak coupling}
approaches have provided a detailed understanding
of the superconducting gap symmetry,
competition of superconductivity with other ordered phases and
occurrence of nematic order,
and traced their common origin to
multi-orbital physics.~\cite{fesc:Yi_orbital_review_nqm_2017}
Fig.~\ref{fig:fesc2} ($c$), from Ref.~\cite{fesc:kuroki_PRB_2009},
shows that the variations
in \tc\ across different 1111 pnictides can be reproduced within RPA
and derive from changes in orbital composition of the Fermi surface.

A first, very elegant, fully first-principles study of superconductivity
in FeSC was carried out by the Gross' group,
who applied their recently-derived {\em ab-initio} theory for spin
fluctuations to FeSe -- see Sect.~\ref{sect:theory:DFT} for details.
Fig.~\ref{fig:fesc2} ($d$) shows the superconducting gap calculated in SCDFT,
which exhibits an $s^{\pm}$ symmetry, consistently with previous {\em hybrid}
studies. It is important to remark, however, that the gap in
Fig.~\ref{fig:fesc2} ($d$) was calculated treating 
the contributions of all
pairing channels (phonons, spin fluctuations, charge fluctuations)
on equal footing, and without any intermediate mapping on an effective
many-body model.
However, even this calculation cannot be considered fully {\em ab-initio}
as, in order to obtain a physically meaningful value for the
superconducting gap and \tc, the authors had to
introduce an artifical {\em scaling} factor in the spin
fluctuation propagator, which would
otherwise diverge.~\cite{DFT:Essenberger_PRB_2014,mine:Ortenzi_PRB_2012}

The divergence is associated to the specific choice of TDDFT
exchange functional (Adiabatic Local Density Approximation, ALDA)
made by the authors which, being the Time-dependent equivalent to
the standard local Density Approximation (LDA),
overestimates the tendency to magnetism in FeSC.~\cite{mine:Mazin_PRB_2008}
Proposals to cure this critical divergence in a non-empirical fashion
are underway,~\cite{DFT:Sharma_sourcefree_jcptc_2018} and could lead to a very important step forward in the understanding of
unconventional superconductors.
However, it is also important to stress that, even if divergence issues
are solved,  \tc\ computed within DFT may still be inaccurate, because
it neglects 
 renormalization effects of quasi-particle energies and
 ineraction vertices due to strong local
 electronic correlations, which, in some FeSC,
 may be sizable.~\cite{DFT:Yin_PRX2011,fesc:demedici_PRL2017}

\section{Outlook and Perspectives}
\label{sect:outlook}

The results described in this chapter show that in the last twenty years there
has been a substantial advancement in the understanding of superconductors,
driven by the progress of {\em ab-initio} electronic structure methods.
The \sh\ discovery has demonstrated that
room-temperature superconductivity can be attained, at least
at extreme pressures,~\cite{H:DrozdovEremets_Nature2015}
and that 
{\em ab-initio} methods
can be very effectively employed to predict new superconductors.
The next challenge  in the field is clearly to devise
practical strategies to replicate the same result at ambient conditions,
exploiting novel synthesis and doping techniques.
Obviously, the development of {\em ab-initio} methods that can treat these
regimes is a crucial step in this direction. 

\begin{center}
\begin{figure}[h!]
\includegraphics[width=12cm,clip,angle=0]{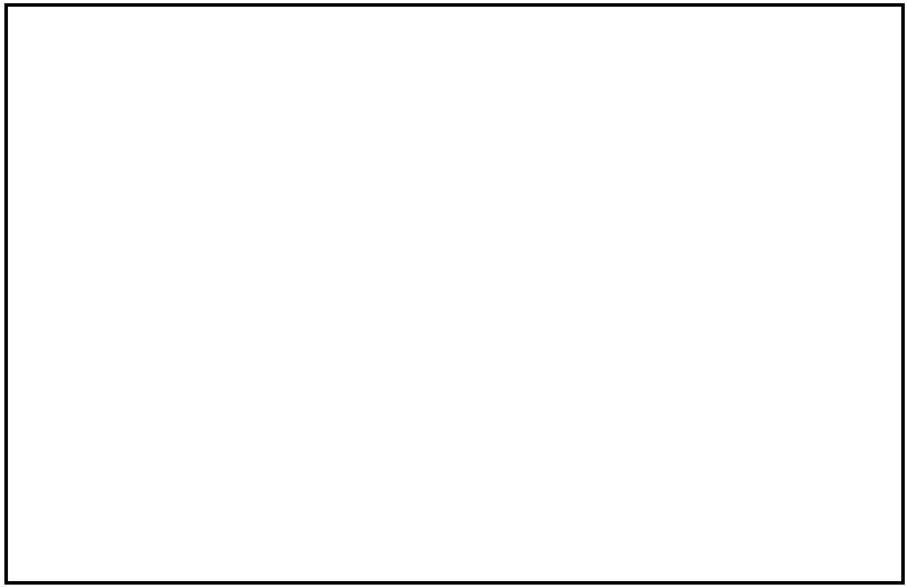}
\caption[]{Selected literature examples of possible strategies to room-temperature superconductivity at ambient pressures:
  ($a$) First-principles prediction of $\alpha$ and $\beta$-LiB~\cite{SC:Kolmogorov_PRB_2006} ({\em Design by Analogy});
  ($b$) Superconductivity in doped ices~\cite{H:Flores_H2O} ({\em Chemical Doping of molecular crystals});
  ($c$) Charge-density profile of superconducting Pb@Si(111) surface.~\cite{SC:Linscheid_PbSi_CM2015} ({\em Atomic-scale design and Dimensionality});
  ($d$) High-pressure superconducting phase diagram of phosphorus~\cite{mine:Flores_PRM2017} ({\em Quenching of High-pressure metastable phases});
  ($e$) \tc\ vs field effect and chemical doping in monolayer MoS$_2$~\cite{SC:Ye_MoS2_gate_science_2012} ({\em Doping by field effect}).
  \label{fig:outlook}}
\end{figure}
\end{center}

The aim of this final section is to give a short overview of promising strategies to high-\tc\ superconductivity, exemplified in Fig.~\ref{fig:outlook}
($a$)-($e$).

\begin{enumerate}[label=\alph*)]
\item {\bf Design by Analogy:}
  This is one of the most common routes to search for new superconductors,
  i.e. achieve high-\tc\ 
  designing materials with a similar geometry, chemistry,
  electronic structure, etc. as the  best existing superconductors.
  This route lead to several predictions of high-\tc\ superconductivity in layered borides and carbides after the discovery of \mg, as well as hole-doped LiBC, and, more recently, doped graphane.~\cite{SC:Satta_beb2_PRB2001,SC:Rosner_LiBC_PRL2002,SC:giustino_PRL_2010}
  The main drawback of many of these early works is that the many of these hypothetical compounds are thermodynamically unstable.
  Nowadays, studies of the thermodynamic stability of compounds are well established, but were very rare at the time of the \mg\ discovery. One of the first studies to take this aspect into account the prediction of superconductivity
  in LiB,~\cite{SC:Kolmogorov_PRB_2006} shown in Fig.~\ref{fig:outlook}($a$).

\item {\bf Chemical Doping of Molecular Crystals:}
  Similarly to covalent solids, molecular ones exhibit stiff bonds, and
  hence their electronic states can couple strongly to lattice vibrations,
  which is a prerequisite for high-\tc\ conventional superconductivity.
  However, similarly to covalent solids, molecular ones are usually
  insulating at ambient pressures.
  Making them superconducting requires either very high pressures,
  or chemical doping, which is often hard to control experimentally and to
  model {\em ab-initio}.

  One of the most complete studies of the effect of doping on superconductivity
  in molecular crystals is the study of superconductivity in doped ice,
  by Flores-Livas et al.~\cite{H:Flores_H2O}
  Here, analyzing the effect of different dopants using supercells,
  the authors found that nitrogen acts as an effective dopant at the oxygen
  site, leading to a \tc\ of 60 K. The study nicely evidences
  the crucial difference between actual doping, modelled with
  supercells, and more approximate approaches, as rigid-band or jellium doping.

\item {\bf Atomic-Scale Design and Dimensionality:}
  Another recent trend in condensed matter research opened by the development of novel techniques, such as Molecular Beam Epitaxy (MBE),
  or exfoliation,~\cite{SC:Novoselov_graphene_Science_2004}
  is the design of materials at the atomic scale, through heterostructuring, controlled deposition, strain engineering, etc.
  This permits to tune superconductivity in existing materials, induce
  it in semi-conductors and semi-metals through doping, like Li-decorated graphene and phosphorene, or create completely artificial superconductors, depositing superconducting elements on semiconductor surfaces, as in Fig.~\ref{fig:outlook}($c$), which shows a SCDFT calculation for Pb on the 111 surface.~\cite{SC:Zhang_PbSi_nphys2010,SC:Linscheid_PbSi_CM2015} For this type of problems, real-space approaches may be more appropriate than reciprocal space ones.~\cite{DFT:Linscheid_PRL_2015}
    
\item {\bf Quenching of High-Pressure Metastable Phases} down to ambient pressure is another very attractive route to stabilize unusual bonding environments, enabled by recent developments in high-pressure techniques.
  In fact, controlled heating and cooling cycles 
  permit to selectively stabilize different metastable phases, realizing
  pressure-hystheresis cycles.
  Figure \ref{fig:outlook} ($d$), from Ref.~\cite{mine:Flores_PRM2017}, shows the crystal structures of the different phases that form the complicated superconducting phase diagram of elemental phosphorus. In this element, a
  high-\tc\ branch, clearly distinct from the low-\tc\ ground-state one, can be accessed by laser heating, and is associated with metastable black phosphorus; similar branchings between high-\tc\ and low-\tc\ phases have been predicted at higher pressures.

\item Doping via {\bf Field Effect} is an attractive route to tune the
  properties of materials continuously, without impurities and distortions associated to chemical doping.  Modern techniques based on liquid electrolytes permit to achieve doping levels as high as $10^{15} carriers/cm^{3}$, corresponding to tenths of electrons, around three orders of magnitude larger than standard solid-state techniques.
  The first succesful applications have been to cuprates, ZrNiCl, and 2D materials --Fig.~\ref{fig:outlook} ($e$).  Rigorous modelling of field-effect devices has been derived in Refs.~\cite{DFT:Sohier_gatingel_PRB2015,DFT:Sohier_gatingph_PRB2017}.
  \end{enumerate}

Although not an experimental technique, methods for machine learning,
which can  be used to pre-screen {\em ab-initio} proposals,
will most likely play a bigger and bigger role in the design
and discovery of new materials.~\cite{DFT:curtarolo_nmat2013}

Note that all of the specific examples discussed above are based on the assumption of {\em conventional} (phonon-mediated) pairing.
The same techniques have been, and can be, applied
also to {\em unconventional} superconductors,
which remain the likeliest candidates for high-\tc\ superconductivity.
However, without a quantitative theory of the superconducting pairing, it is at the moment impossible to formulate reliable predictions of \tc\ and other superconducting properties.  A progress in this direction is therefore an essential prerequisite to any meaningful computational search.



\begin{acknowledgement}

There are many people who, over the years, helped me to shape my view
on superconductivity. Many of these encounters turned into friendships,
and I am very grateful for that.
A special thank goes to my mentors in Rome (Luciano Pietronero,
Giovanni Bachelet) and Stuttgart (Jens Kortus and Ole Krogh Andersen),
who introduced me to the field of superconductivity and electronic structure,
 as well as to all my collaborators and students, with whom I had the pleasure to work and argue on many topics.
Thanks to 
Jos\'e Flores-Livas, Christoph Heil,
Renato Gonnelli, Bernhard Keimer,
Jun Sung Kim, Rheinhard Kremer, Igor Mazin, Paolo Postorino,
Gianni Profeta abd Antonio Sanna
for the many discussions and projects we shared over the years.

I would never have completed this chapter without the help of my current
office neighbor, Paolo Dore, who enquired about the status of the
project almost every day, and of Luca de' Medici, Christoph Heil,
Antonio Sanna and Alessandro Toschi, who gave me suggestion on parts
of the manuscript at different stages.
Finally, I would like to dedicate this work to the memory of
two very special people, Sandro Massidda and Ove Jepsen, that I will
always remember for their kindness, culture, and enthusiasm for physics.
I miss them both.

\end{acknowledgement}

%
%
%

%
%
%

%

%

\end{document}